%% file: AN-Homeo.tex
\documentclass[aps,pre,reprint,superscriptaddress]{revtex4-2}

\usepackage{graphicx}
\usepackage{subfigure}
\usepackage{color}
\usepackage[dvipsnames]{xcolor}
\usepackage{amsmath,amssymb}
\usepackage[linktocpage,colorlinks=true,linkcolor=blue,citecolor=blue,breaklinks=true,urlcolor=blue]{hyperref}
\usepackage{verbatim}
\usepackage{breakcites}
\usepackage{wrapfig}
\usepackage{soul}
\usepackage{xparse}
\usepackage{multirow}
\usepackage{bm}
\usepackage[shortlabels]{enumitem}
\usepackage{graphicx}

\input{custom_commands.tex}

\input{crossref_helper.tex}
\myexternaldocument{SI}

\newcounter{siequation}
\newcounter{sifigure}
\newcounter{sitable}
\newcounter{sisection}
\newcounter{simovie}
\begin{document}


\title{Topological kicks enhance colloidal diffusivity in topological turbulence}

\author{Timofey Kozhukhov}
\affiliation{\afilSopa}
\affiliation{\afilNbi}

\author{Benjamin Loewe}
\affiliation{\afilChile}

\author{Kristian Thijssen}
\affiliation{\afilNbi}

\author{Tyler N. Shendruk}
\affiliation{\afilSopa}

\date{\today}

\begin{abstract}
\noindent
Colloidal inclusions in nematic fluids induce topological defects that govern their dynamics. 
These defects create  well-understood rheological behavior in passive nematics, but the interplay between colloid-associated defects and spontaneously generated activity-induced defects introduces new dynamical regimes in active nematic turbulence. 
Using mesoscale simulations, we study the motion of colloids with strong anchoring in an active nematic and find effective colloid diffusivity exhibits a striking non-monotonic dependence on activity.  
At low activities, \emph{topological kicks} from the motile activity-induced bulk defects drive frequent rearrangements of the colloid-associated defects, enhancing colloidal transport. 
At high activity, defect interactions become isotropic, decorrelating colloidal motion and reducing the effective diffusivity. 
Our results reveal how the competition between colloid-associated and activity-induced defects fundamentally shapes transport in active nematics.
\end{abstract}



\maketitle


\begin{figure*}
    \centering
    \includegraphics[width=0.95\linewidth]{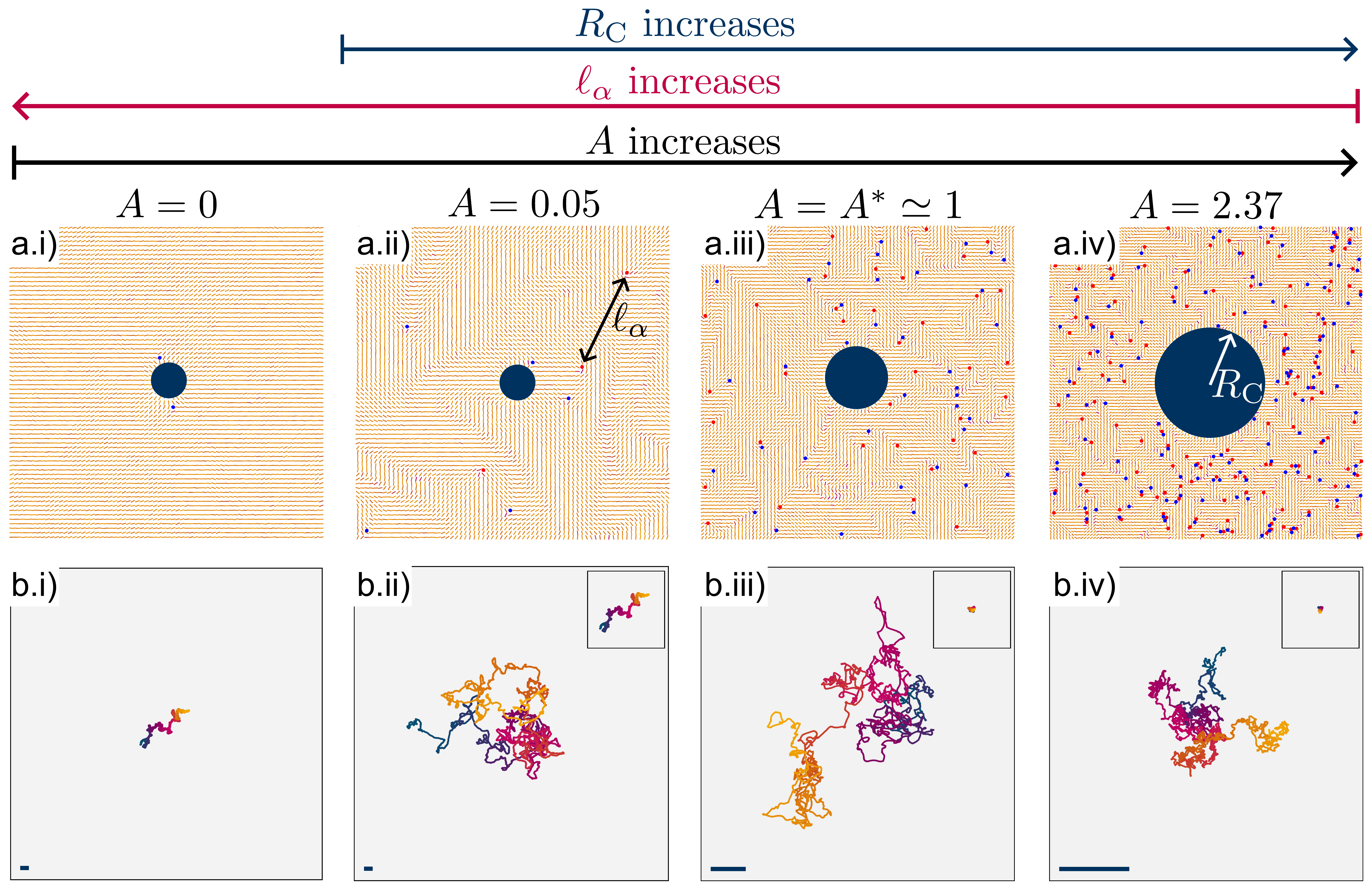}
    \bcaption{
        Topological turbulence enhances displacement of strongly anchored colloids.
    }{
        \textbf{(a)} 
        Director field snapshots for varying activity number, $\actNum$.
        Red and blue dots denote $+1/2$ and $-1/2$ defects, respectively. 
        \textit{i.} $A=0$
        \textit{ii.} $0.27$
        \textit{iii.} $0.67$
        \textit{iv.} $2.37$
        \textbf{(b)}
        Colloid trajectories over $3\times10^5$ timesteps. 
        Each trajectory colored by time and each domain covers the same area of $\lenSys^2=160^2$, with scale bars indicating the colloid radius. 
        Insets show a passive trajectory of a colloid with identical radius.
    }
    \label{fig:fig1A}
\end{figure*}

Many intrinsically out-of-equilibrium biological materials have recently been shown to exhibit nematic symmetry~\cite{marchetti2013hydrodynamics}. 
Examples include subcellular filaments~\cite{sanchez2012spontaneous, zhang2018interplay, maroudas2021topological}, bacterial biofilms~\cite{dell2018growing, van2024emergent} and cell monolayers~\cite{Duclos2014, copenhagen2021topological, duclos2017topological, ruider2024topological}.
These typically do not exhibit global symmetry breaking, but rather include local disruptions to the orientational order, known as topological defects, which are increasingly recognized as key to determining biomaterial properties~\cite{senyuk2012shape, ohzono2012zigzag, kosterlitz2016topological, bowick2022symmetry, ruider2024topological, ardavseva2022topological}, just as they are in passive materials~\cite{Kralj2018}.
Defects act as biological hotspots, influencing apoptosis~\cite{Saw2017}, neural mound formation~\cite{kawaguchi2017topological}, and limb development in Hydra~\cite{maroudas2021topological}. 
For this reason, they offer a pathway toward designing life-like synthetic materials~\cite{senyuk2012shape,ohzono2012zigzag}.  

Defects can be induced in passive liquid crystals by strong anchoring to boundaries, as occurs when colloids are suspended in nematics~\cite{tkalec2013topology}. 
In three dimensions, they form entangled disclination lines~\cite{head2024}, while they form tightly bound colloid-defect pair complexes~\cite{tkalec2013topology}.  
Defects interact dynamically with their environments, yielding intriguing properties, such as disclination knotting in nematic gels~\cite{vcopar2015knot}, lock-key control~\cite{Head2024-WavyWalls}, optical trapping~\cite{senyuk2012shape} and tunable three-dimensional architecture~\cite{modin2023tunable}.

Another strategy for creating defects is through energy injection, via temperature gradients, shearing or local biological activity. 
In active nematics, local energy input is converted into mechanical work, continuously driving defect nucleation~\cite{shankar2018defect}, which maintains a non-zero steady-state defect density~\cite{thampi2013velocity,doostmohammadi2018active}. 
Active nematic defects exhibit unique non-equilibrium properties, such as self-propulsion~\cite{giomi2014defect}, which can profoundly influence collective dynamics~\cite{shankar2019hydrodynamics} and lead to active---or topological---turbulence~\cite{Alert2022AnnRevCondMat}. 

Despite extensive studies of both active nematics and colloidal systems, the interplay between colloid-induced surface defects and bulk defects in active fluids remains uninvestigated. 
Studies of geometry-induced dynamics, such as active nematics on the surface of a toroid~\cite{ellis2018curvature}, confined to boundaries~\cite{hardouin2022active} or within emulsions~\cite{guillamat2018active,negro2025}, have revealed rich and complex behaviors.
Similarly, early studies on colloidal solutes in active solvents have shown activity alters effective transport dynamics~\cite{valeriani2011colloids,lagarde2020colloidal,Loewe2021NJP,Alexander2023} and interactions with boundaries~\cite{Neville2024}. 
However, few studies have explored systems in which active stresses sustain a steady-state population of defects that interact with colloid-induced surface defects. 
This represents an unexplored avenue for potentially anomalous behaviors. 

In this manuscript, we investigate how the active topological environment surrounding strongly anchored normal homeotropic colloids influences the dynamics in active nematic topological turbulence.
We uncover striking results for the effective diffusivity of colloids, revealing a non-monotonic dependence on dimensionless activity, which arises due to interactions between colloid-associated companion defects and activity-induced bulk defects. 

The coarse-grained mesoscale simulation technique Multi-Particle Collision Dynamics (MPCD; \S~SI) is used to model the orientational fluid in the limit of a one-viscosity and one-elastic constant approximation, $K$~\cite{Shendruk2015SoftMatter-NMPCD}. 
This approach has been employed to simulate polymer-nematic mixtures~\cite{valei2025-PassivePolymerMPCD} and colloidal liquid crystals~\cite{head2024}. 
A strongly anchored colloid in a passive nematic induces a pair of nearby $-1/2$ defects (\fig{fig:fig1A}a.i).
We refer to the colloid-associated defects as \emph{companion defects} and the structure as a \emph{colloid-companion complex}. 
In a passive nematic, the colloid-companion complex has a neutral topological charge. 
The nematic MPCD fluid is given \emph{extensile} activity by including a local active stress $\act \geq 0$, resulting in an active length scale $\lenAct \sim \left(K/\act\right)^{1/2}$ corresponding to the average spacing between nematic defects~\cite{Hemingway2016SoftMatter}.
The active length scale can be compared to colloid radius through the dimensionless activity number
\begin{equation}\label{eq:actNum}
    \actNum = \left(\frac{\colRad}{\lenAct}\right)^2 \left( 1 + 2\frac{\colRad}{\lenAct} \right)^{-1}. 
\end{equation}
Specifically, $\actNum$ compares the colloidal area $\pi\colRad^2$ to the area of an annulus of width $\lenAct$ around the colloid $\pi \left( \lenAct^2 + 2\colRad\lenAct \right)$.  
When $\actNum>1$, the colloid is large compared to the intrinsic length scale of the topological turbulence, whereas the colloid is small compared to the distance between activity-induced defects when $\actNum<1$. 


When activity is sufficient to produce topological turbulence, the active stress induces the continuous nucleation of defects in bulk away from the colloid. 
For low activity, one of the two companion defects is frequently lost from the colloid-companion complex --- only a single $-1/2$ companion defect remains bound to the colloid (\fig{fig:fig1A}a.ii, Movie~M1). 
This structure persists to intermediate activities (\fig{fig:fig1A}a.iii, Movie~M2), but, at higher activities, a sea of topological defects surrounds the colloid, such that the notion of well-defined companion defects becomes ill-defined  (\fig{fig:fig1A}a.iv, Movie~M3). 
This change in topological configuration will be shown to have profound effects on the dynamics (\fig{fig:fig1A}b). 

\begin{figure*}
    \centering
    \includegraphics[width=0.99\linewidth]{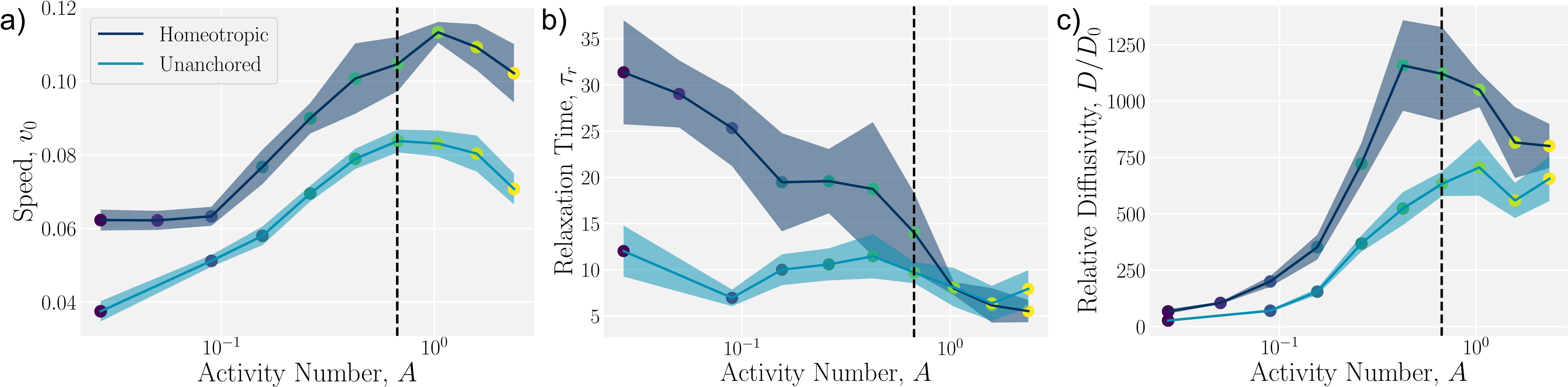}
    \bcaption{
          Mean-squared displacement (MSD) dynamics of passive colloids in topological turbulence. 
    }{
        Measured fit values from MSDs.
        Shaded regions represents the standard deviation between realizations, and the dashed line represents $\actNum=\actCrit$.
        \textbf{(a)} 
        Speed $v_0$. 
        \textbf{(b)} 
        Decorrelation time $\tau_\mathrm{r}$.
        \textbf{(c)}
        Relative diffusivity $\Deff/\DBrownian$. 
        Strongly anchored homeotropic colloids exhibit non-monotonicity, whereas unanchored colloids do not. 
    }
    \label{fig:fig2rework}
\end{figure*}

Example trajectories illustrate that the displacement of the colloid-companion complex increases as activity increases (\fig{fig:fig1A}b.i-ii) to a maximum (\fig{fig:fig1A}b.iii) but reduces once the colloid loses both companion defects (\fig{fig:fig1A}b.iv). 
In all cases, the active systems exhibit enhanced diffusivity compared to an identically sized colloid in a passive nematic (\fig{fig:fig1A}b insets).
This difference is quantified by the mean-squared displacement (MSD, \S~II; Fig.~S1) of the colloid trajectories. 
The instantaneous speed $\colSpeed$ of the colloid obtained from the MSD increases with activity $\actNum$ (\fig{fig:fig2rework}a) and plateaus as activity increases above a value of $\actCrit\simeq 0.67$. 
Simultaneously, the decorrelation time $\tau_\mathrm{r}$ of the colloid monotonically decreases with $\actNum$ (\fig{fig:fig2rework}b). 
These changes in relaxation time and speed result in a peak in the effective diffusivity $D$ relative to its thermal counterpart $\DBrownian$ (Fig.~S2) around a critical activity $\actCrit = 0.67$ (\fig{fig:fig2rework}c).
This reveals that not only does the active component of diffusivity $v_0^2\tau_r$ dominate over the passive thermal component, but is maximized when the colloidal size equals the active length scale, \ie $\actNum=\actCrit \simeq 1$. 
The dynamics uncovered here of the strongly anchored colloid are unlike the behaviors of unanchored colloids, which lack companion defects. 
While unanchored colloids exhibit monotonically increasing speed (\fig{fig:fig2rework}a), their decorrelation time remains small and nearly constant (\fig{fig:fig2rework}b; \S~SIII). 
The unanchored colloids exhibit enhanced diffusivity, as is to be expected since the active flows increase dispersion, but do not exhibit a maximum, indicating that the maximum in effective diffusivity is due to the interplay between companion defects and the topological turbulence. 

\begin{figure}
    \centering
    \includegraphics[width=0.85\linewidth]{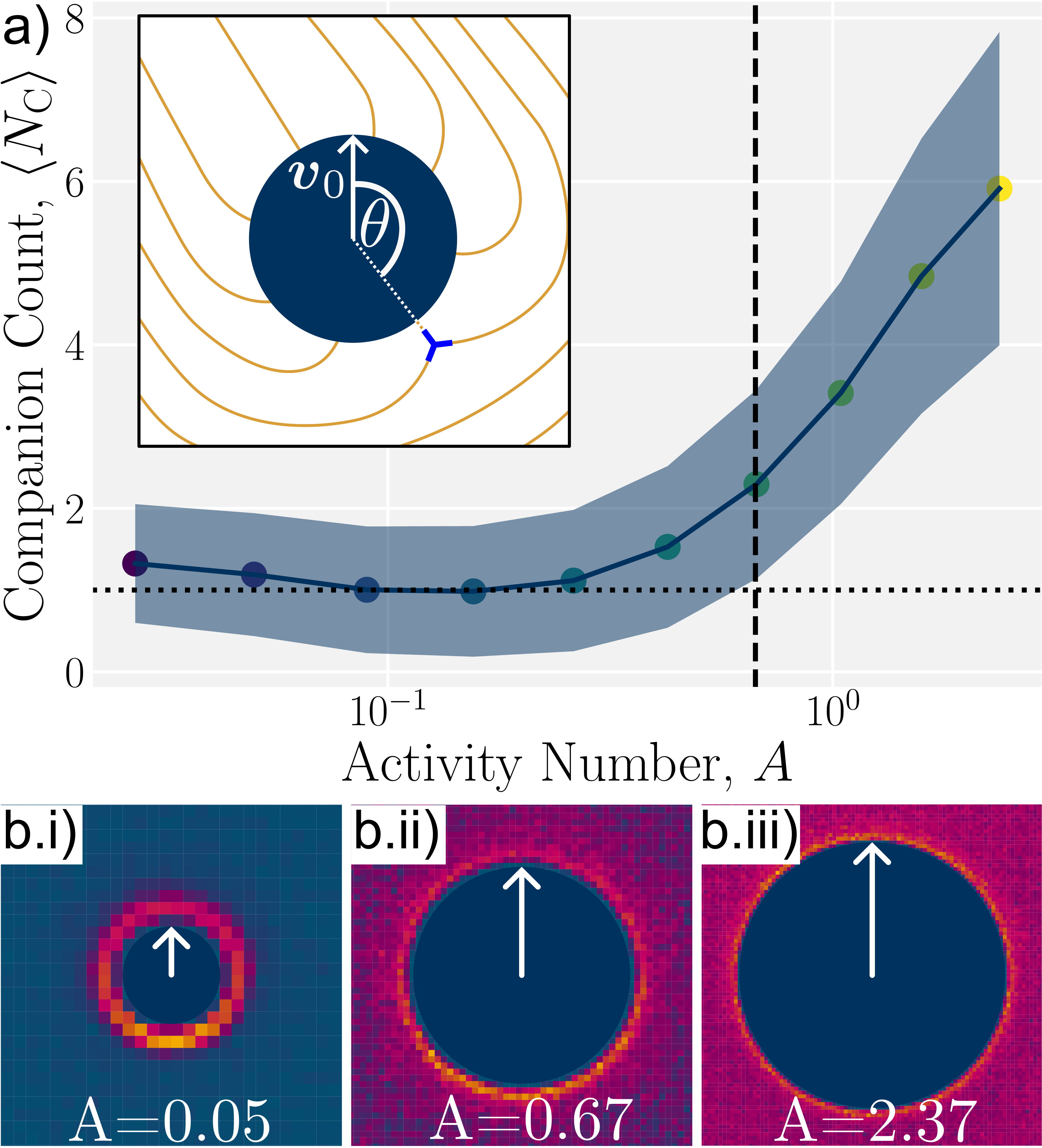}
    \bcaption{
        Radial symmetry breaking of $-1/2$ defects.
    }{
        \textbf{(a)}
        Average number of $-1/2$ defects $N_\mathrm{C}$ in the near colloid region.
        Shaded region represents standard deviation between realizations, dashed line represents $\actNum=\actCrit$ and dotted line represents a single companion defect bound to the colloid. 
        \textbf{(inset)}
        Schematic of defect angle $\theta$ with respect to colloid velocity $\colVel$.
        Complexes with a single companion form a far-field equivalent a $+1/2$ defect, oriented towards the companion.
        \textbf{(b)}
        Distribution of $-1/2$ defects in the moving colloid reference frame (white arrow).
        Blue represents low probability and yellow high. 
        Distributions for all considered activity numbers are shown in Fig.~S6. 
    }
    \label{fig:fig3rework}
\end{figure}

The colloidal dynamics are due to changes to the colloid-companion complex structure. 
For low activities, the $-1/2$ defect distribution is highly localized around the colloid, with increases to the activity homogenizing this distribution (Fig.~S3a; \S~SIV).
However, the average distance of the closest $-1/2$ defect to the colloid decreases with activity (Fig.~S3b-c), which allows the definition of a \emph{near-colloid region} within which the number of $-1/2$ defects $N_\mathrm{C}$ can be counted (\fig{fig:fig3rework}a). 
At low activity, $N_\mathrm{C}\approx1$ with a high probability of finding a single defect (Fig.~S4).
At higher activities, $N_\mathrm{C}$ rises steadily, and the structure of the surrounding defects becomes more gas-like.

The observation of a single companion defect at low activities is at odds with the expectation from passive nematics~\cite{tkalec2013topology}, suggesting the loss of one companion $-1/2$ defect is facilitated by the topological turbulence. 
Self-propelled $+1/2$ defects in the surrounding turbulence are continuously created and annihilated, with annihilation events occurring because of the attraction between oppositely charged defects~\cite{Giomi2014PhilTransactionsA,Thampi2014EPL}.
Indeed, as an activity-induced $+1/2$ defect from the bulk turbulence approaches the colloid-companion complex, it modifies the elastic energy landscape. 
A simplified model of the elastic interactions between a colloid-companion complex with a colloid of $+1$ charge, two antipodal $-1/2$ companion defects and an approaching activity-induced $+1/2$ defect from the surrounding steady-state population of defects (\S~SV; Fig.~S5a) shows that the energy barrier for the bound companion defect is large when the $+1/2$ defect is far from the complex (Fig.~S5b). 
However, when the $+1/2$ defect is sufficiently close, the potential barrier of the nearest $-1/2$ shrinks to zero, allowing it to escape into the bulk turbulence. 
Moreover, the loss of one companion defect enhances the elastic binding between the remaining companion defect and the colloid (\S~SV; Fig.~S5c). 
Thus, the remaining colloid-companion complex with only a single bound defect is more stable and maintains a non-neutral charge of $+1/2$, which further repels $+1/2$ defects.

Since the colloid-companion complex with a single bound $-1/2$ defect has a net $+1/2$ charge, it possesses orientational symmetry breaking (\fig{fig:fig3rework}a inset). 
In the colloidal reference frame (\fig{fig:fig3rework}b, Fig.~S6), $-1/2$ defects are not uniformly distributed around the colloid at low activities (\fig{fig:fig3rework}b.i). 
Instead, defects have a higher probability of being found trailing the colloid's direction of motion. 
This anisotropy persists as the activity approaches $\actCrit$ (\fig{fig:fig3rework}b.ii), but the distribution becomes radially symmetric at the highest activities (\fig{fig:fig3rework}b.iii), as quantified by distributions of relative angles (\S~SVI; Fig.~S7).

Since the structure of the single-companion complex in the far-field limit (is equivalent to a $+1/2$ defect (\fig{fig:fig3rework}a inset), one might suppose that the net $+1/2$ topological charge of the colloid-companion complex endows it with the self-motility of an effective $+1/2$ defect, which would cause it to move away from the primary splayed region towards the primarily bend region in front in extensile active fluids~\cite{Head2024-DTensor}. 
Surprisingly, this effective $+1/2$ far-field director pattern would imply motion in the opposite direction to the observed colloidal motion.
Thus, the enhanced diffusion is not directly due to the effective $+1/2$ topological structure of the colloid-companion complex, but due to a secondary effect.

\begin{figure}
    \centering
    \includegraphics[width=0.85\linewidth]{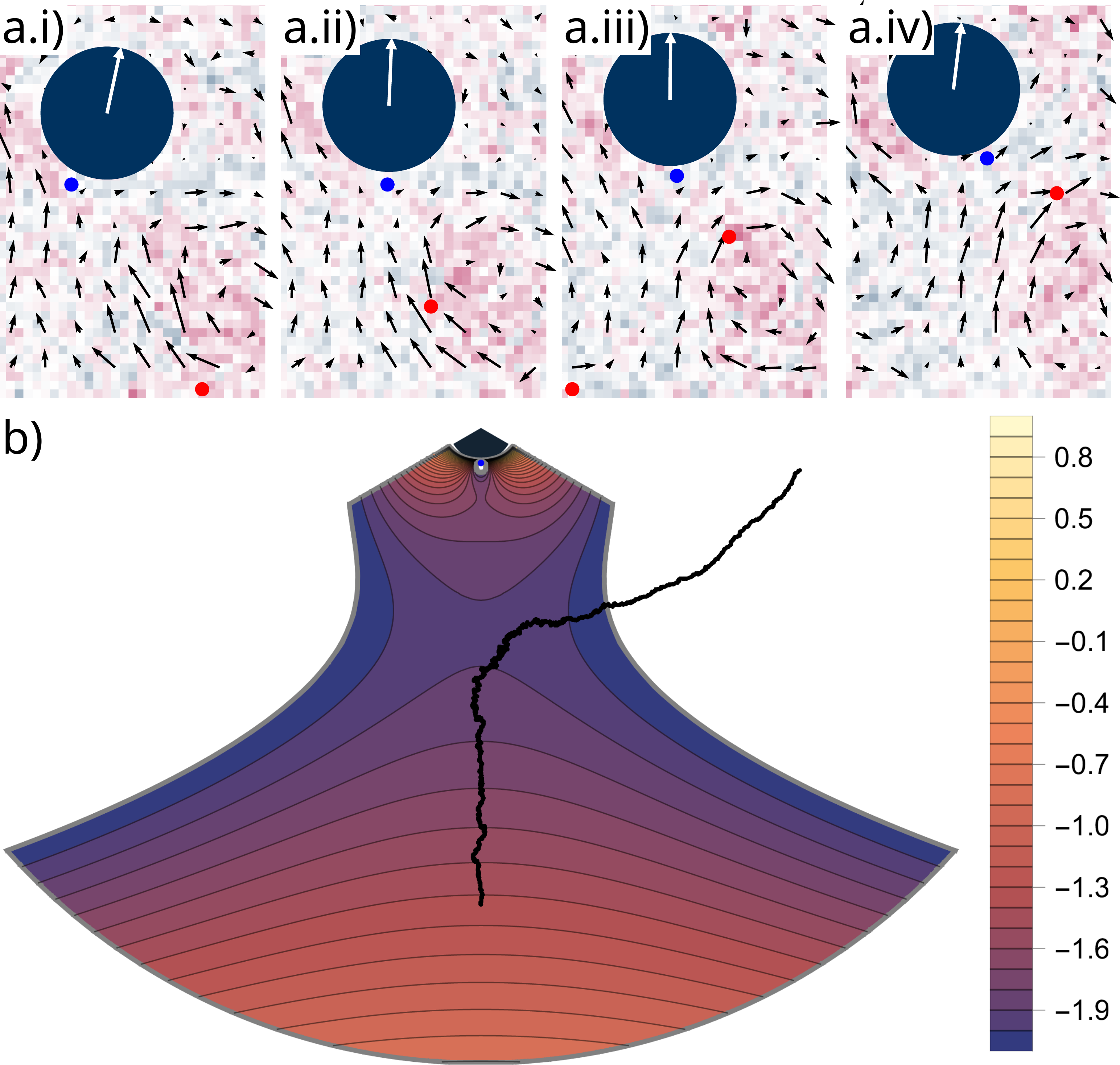}
    \bcaption{
        Attempted annihilation of a $-1/2$ companion defect by and activity-induced $+1/2$ defect results in enhanced colloidal diffusivity.
    }{
        \textbf{(a)}
        Simulation snapshots of velocity (black arrows) and vorticity (color map) at the critical activity $\actCrit\simeq0.67$.
        Snapshots are cropped and rotated from a fixed space within the simulation domain, visible in Movie~M4. 
        The white arrow indicates instantaneous colloid direction of motion and defects marked as in \fig{fig:fig1A}. 
        \textit{a.i)}
        Time $t=1680\dt$:
        \textit{a.ii)}
        $1780\dt$:
        \textit{a.iii)}
        $1880\dt$:
        \textit{a.iv)}
        $2010\dt$:
        \textbf{(b)}
        The potential $U_\mathrm{eff}$ experienced by a $+1/2$ defect approaching a colloid-defect complex (Eq.~S9). 
        Example Langevin trajectory of a $+1/2$ defect plotted as a black line, illustrating typical deflection dynamics.
    }
    \label{fig:fig4rework}
\end{figure}

The colloid-companion complex moves in the opposite direction due to the interplay between the colloid-companion complex and activity-induced $+1/2$ defects from the surrounding topological turbulence (\fig{fig:fig4rework}a). 
The incoming $+1/2$ defects are predominantly found in the direction opposite to the direction of the colloid velocity at low activities (\S~SVII; \fig{fig:fig4rework}a.i). 
This is because they are attracted to the remaining $-1/2$ companion defect.
As an activity-induced $+1/2$ defect approaches the companion, it experiences an elastic attraction to the companion $-1/2$ defect, while also producing an active flow (\fig{fig:fig4rework}a.ii)~\cite{giomi2014defect}. 
The flows generated by the approaching $+1/2$ defect advect the colloid, pushing it away from the approaching $+1/2$ and companion $-1/2$ defects (\fig{fig:fig4rework}a.iii).
However, rather than annihilate with the companion, the $+1/2$ defect performs a \emph{topological kick} to the colloid as it is deflected (\fig{fig:fig4rework}a.iv).

The effective energy landscape experienced by the $+1/2$ defect as it approaches the complex explains its deflection, rather than annihilation (\S~SVIII). 
The effective potential is obtained by considering the elastic energy with the colloid-companion complex and a linear potential that models a constant active self-propulsive force. 
Crucially, the colloidal monopole and the dipole center are off-set, with the dipole laying closer to the $+1/2$ defect. 
The resulting effective potential due to elastic interactions, anchoring, and self-motility is saddle-like in nature (\fig{fig:fig4rework}b; Eq.~S9), which is illustrated by a typical Langevin trajectory taken by an activity-induced $+1/2$ defect as it approaches the complex (\fig{fig:fig4rework}b). 

It is by this very mechanism that the colloid-companion complex is kicked through the topological turbulence with enhanced diffusion:
$+1/2$ defects are attracted to the companion defect of the colloid, seeking to annihilate, but are instead deflected by the elastic forces from colloidal anchoring. 
The flow fields produced by the self-motile $+1/2$ defect push the colloid-companion complex away, in the direction opposite the $-1/2$ companion defect, resulting in enhanced colloidal diffusivity.
This is the opposite direction as would be expected due to the effective $+1/2$ topological charge of the colloid-companion complex (\fig{fig:fig3rework}a inset).
At low activity numbers, the lower number of $+1/2$ defects in the fluid bulk causes fewer of these events to occur. 
Thus, the colloid-companion complex is slower (\fig{fig:fig2rework}a) but the persistence time is longer $\tau_\mathrm{r}$ (\fig{fig:fig2rework}b).
As the activity number increases, the number of $+1/2$ defects in the fluid bulk increases, leading to more near-annihilation events, with an increase in the instantaneous colloidal motion but decrease in persistence time and a loss of anisotropy. 
In this way, the model also explains the non-monotonic diffusivity observed in strongly anchored homeotropic colloids, but not in unanchored colloids. 


There is a rich history of research investigating the dynamics of colloidal particles in passive nematics, showing how topological interactions arising through anchoring can lead to the formation of defect structures~\cite{head2024} and colloidal self-assembly~\cite{Musevic2006,Skarabot2008}. 
Activity restructures the topological environment of the colloid from a neutral colloid-companion complex to a positively charged colloid-companion complex to a particle in a gas of activity-induced defects. 
Our results begin to extend this understanding to active nematics, where spontaneous flows and self-propelled defects have already been shown to rectified motion in chiral inclusions~\cite{Sattvic2023,Alexander2023}, enhance mixing properties of tracers above active carpets~\cite{Guzman2021}, and generate complex motion in emulsions~\cite{negro2025}. 
The introduction of anchoring in active nematics, as explored here, reveals a novel mechanism by which the colloid-defect interactions are fundamentally altered, leading to enhanced colloidal diffusivity and defect-mediated propulsion arising from topological kicks.

This study focused on strong normal anchoring at the surface, but planar anchoring is also common in the passive nematic colloidal literature. 
The resulting colloid also has an effective $+1$ topological charge~\cite{tkalec2013topology}, and may similarly lead to enhanced diffusivity.
While this work has primarily focused on how active nematics influence colloidal dynamics, it also reveals how colloidal inclusions---through anchoring and defect interactions---reciprocally influence the structure and flow of the active nematic medium. 
This dual interaction opens the door to using colloids as a means of manipulating and controlling defect behavior in active systems, ultimately paving the way for the exploration of colloidal self-assembly in topological turbulence as an emergent and tunable phenomenon.

\vspace{\baselineskip}

\paragraph*{Acknowledgments}
We acknowledge useful discussions with Davide Marenduzzo.
This research has received funding from the European Research Council (ERC) under the European Union’s Horizon 2020 research and innovation programme (Grant agreement No. 851196). 
This work was further supported by the Novo Nordisk Foundation grants no. NNF18SA0035142 and NNF23OC0085012.
For the purpose of open access, the author has applied a Creative Commons Attribution (CC BY) license to any Author Accepted Manuscript version arising from this submission.

    
    

\bibliography{biblio.bib}

\end{document}


\title{Supplementary Material: \\ Topological kicks enhance colloidal diffusivity in topological turbulence}

\author{Timofey Kozhukhov}
\affiliation{\afilSopa}
\affiliation{\afilNbi}

\author{Benjamin Loewe}
\affiliation{\afilChile}

\author{Kristian Thijssen}
\affiliation{\afilNbi}

\author{Tyler N. Shendruk}
\affiliation{\afilSopa}

\date{\today}

\maketitle

\section*{Movie Captions}

\begin{enumerate}[label=M\arabic*, leftmargin=1em]
    \item \label{mov:dir-lowA}
        \textbf{Anchored colloidal particles in active turbulence exhibit enhanced motility.}
        Director field, where each line segment corresponds to the director $\dir_\cell$ averaged across four MPCD cells, colored by scalar order parameter $S_C$. 
        Simulation parameters are $\actNum = 0.05$ and $\lenSys=70$.
        Common simulation parameters with \movie{mov:dir-midA} and \movie{mov:dir-highA} include a simulation time length of $20000 \dt$ following a $2000 \dt$ warmup, each frame is $10\dt$ apart, and the mean field potential is $\mfp=10$.
        $+1/2$ and $-1/2$ defects are marked as red and blue circles respectively.
    \item \label{mov:dir-midA}
        \textbf{Increasing activity number $\actNum$ to a critical value maximizes colloid motility.}
        Director field $\dir_\cell$ averaged across four MPCD cells, colored by scalar order parameter $\cellScalarOrder$. 
        Simulation parameters are $\actNum = 0.67$ and $\lenSys=174$.
        Uses the same common simulation parameters as \movie{mov:dir-lowA}.
    \item \label{mov:dir-highA}
        \textbf{Larger activity numbers $\actNum$ lower colloidal diffusion and saturate colloidal ballistic speeds.}
        Director field $\dir_\cell$ averaged across four MPCD cells, colored by scalar order parameter $\cellScalarOrder$. 
        Simulation parameters are $\actNum = 2.37$ and $\lenSys=200$.
        Uses the same common simulation parameters as \movie{mov:dir-lowA}.
    \item \label{mov:DeflectionMov}
        \textbf{Attempted annihilation of a companion defect results in enhanced motility of the colloid.}
        Velocity field $\vec{v}$ (black arrows) colored underneath by scalar value of vorticity $\omega$.
        $\actNum=0.67\simeq\actCrit$, $\lenSys=76$, the corresponding simulation length is $350\dt$, and each frame is $10\dt$ time units apart.
\end{enumerate}

\section*{Supplementary Text}

\subsection{Numerical approach}\label{si:sec:methods}

\subsubsection{Solvent model}\label{si:sec:solvent-model}

The solvent is modeled by the particle-based MPCD technique~\cite{GompperIhle2009Book-MPCD}, in which particle interactions and momentum exchange are encoded through lattice-based momentum conserving collision operators. 
The collision operators control the material properties of the solvent being simulated, while rigid-body dynamics between the off-lattice fluid particles and colloid surfaces resolve suspended colloid dynamics (see \sect{si:sec:solute-model}).
Solvent particle $i$ is treated as a point particle with mass $\mass$, continuous position $\vec r_i$ and velocity $\vec v_i$, which undergoes a ballistic streaming step
\begin{equation}
    \vec r_i(t+\dt) = \vec r_i(t) + \vec v_i(t)\dt,
\end{equation}
over the timestep size $\dt$.
Solvent particles are then binned into lattice-based cells of size $\cellLen$, and undergo a collision operation in which particles within each cell $\cell$ undergo a stochastic collision operation that conserves mass and momentum within each cell
\begin{equation}
    \vec v_i(t + \dt) = \vcm_\cell(t) + \colop_{i, \cell}(t),
\end{equation}
where $\vcm_\cell$ is the center of mass velocity of the cell, and $\colop_{i, \cell}$ is the collision operator acting on particle $i$ within cell $\cell$. 
All $\mathcal{N}_\cell$ particles within the cell stochastically exchange momentum. 

The MPCD collision operator encapsulates all solvent dynamics through particle interactions, making it possible to simulate different types of fluids by varying the operator~\cite{Gompper2008JCP-ViscoelasticMPCD, Toneian2019JCP-ViscoelasticMPCD, Kapral2013-ReactiveMPCD, Shendruk2015SoftMatter-NMPCD}. These operators are cumulative, enabling the design of new collision models, which led to the development of an active nematic collision operator (AN-MPCD), which successfully reproduces active nemato-hydrodynamics across a selected range of activity parameters $\act$.

AN-MPCD is built upon a collision operator for isotropic fluids with an integrated Andersen thermostat, $\colAndersenBasic$~\cite{Gompper2007EPL}. 
The Andersen operator adds random Gaussian velocities to the center of mass velocity of the cell for each particle $i$ in cell $\cell$, sampled from a Maxwell-Boltzmann distribution at thermal energy scale $\kbt$. To conserve momentum, the average Gaussian velocity within each cell is subtracted from the sampled velocities, ensuring zero change in momentum per cell while maintaining a uniform temperature across the system.
By similarly subtracting the residual cellular angular momentum, the operator can be extended to conserve angular momentum, yielding the angular-momentum-conserving Andersen thermostat, $\colAndersen$~\cite{GompperIhle2009Book-MPCD}.

Conserving angular momentum allows for the creation of a nematic collision operator for passive nematic fluids $\colNem$~\cite{Shendruk2015SoftMatter-NMPCD}:
By treating the MPCD particles as point-particle rods rotating in a background fluid, with a nematic orientation $\ori_i$ assigned to each particle, a coarse-grained cellular nematic director $\dir_\cell$ and scalar order parameter $\cellScalarOrder$ are obtained.
An \emph{orientation} collision operator is applied to the particle orientations within each cell through a local mean field model of the nematic fluid, producing local variations in orientation~\cite{Shendruk2015SoftMatter-NMPCD}
\begin{equation}
    \ori_i(t + \dt) = \vec\Psi_{i,\cell} \left(
        U, \tens{Q}_\cell, \ori_i(t)
    \right).
\end{equation}
The orientation collision operator $\vec\Psi_{i,\cell}$ performs a rotational perturbation about the local nematic director $\dir_\cell$ through an angle $\theta_i$ obtained by sampling the Maier-Saupe distribution~\cite{Shendruk2015SoftMatter-NMPCD}.
The degree of local ordering can be controlled through a mean-field parameter $\mfp$, which controls the isotropic-nematic transition and can be mapped to the Frank elastic constant $K$~\cite{Shendruk2015SoftMatter-NMPCD}.
This orientation field has two-way coupling to the velocity field. 
The orientation is coupled to velocity gradients through a Jeffery torque applied to the particle orientations. 
The velocity is coupled to the rate of change of orientation through angular momentum conservation via the nematic collision operator $\colNem$, which simulates backflow.
The tumbling behavior of nematogen rods is governed by the tumbling parameter $\lambda$ and a shear coupling coefficient $\chi_\mathrm{HI}$, which adjusts the alignment relaxation time relative to $\dt$~\cite{Shendruk2015SoftMatter-NMPCD}.

To reproduce active nemato-hydrodynamics, an out of equilibrium collision term is appended to the nematic collision operator, resulting in a collision operator for active nematic fluids $\colAN$ (AN-MPCD)~\cite{Kozhukhov2022-ANMPCD}.
This term represents a dipolar force to the particles within each cell, applying impulses of magnitude $\act_\cell$ to the particles away from the center of mass of the cell in the direction of the local nematic director $\pm \dir_\cell$~\cite{Kozhukhov2022-ANMPCD}.
To mitigate the effects of density-driven fluxes in the solvent, the local activity $\act_\cell$ is modulated by a local density dependent kernel, giving \emph{modulated cell-carried activity}~\cite{Kozhukhov2024-ANSig}, which are further controlled by parameters $\Sigwid$ and $\Sigpos$ which tune the degree of modulation.

\subsubsection{Solute model}\label{si:sec:solute-model}

A colloidal disk of radius $\colRad$ immersed within the MPCD solvent is simulated as a free impermeable surface that interacts with the solvent particles through rigid-body collisions~\cite{head2024}.
Solvent MPCD particles are detected to collide with the circular boundary surface, upon which a set of deterministic collision rules are applied
\begin{align}
    \vel_i &\to M_{\vel, \mathrm{N}} (\vel_i \cdot \norm) \norm  + M_{\vel, \mathrm{T}} (\vel_i \cdot \tang) \tang = \vel_i^\dag\label{eq:mpcd-bc-vel-rule}\\
    \ori_i &\to M_{\ori, \mathrm{N}} (\ori_i \cdot \norm) \norm + M_{\ori, \mathrm{T}} (\ori_i \cdot \tang) \tang = \ori_i^\dag, \label{eq:mpcd-bc-ori-rule}
\end{align}
where $\norm$ and $\tang$ are the normal and tangential vectors to the boundary surface at the point of collision, respectively.

\begin{figure}[tb]
    \centering
    \includegraphics[width=0.7\linewidth]{Figures/SI_Misc/msd_fits.pdf}
    \bcaption{Mean-square displacements for anchored colloids exhibit different behavior with $\actNum$}{
        MSD ($\av{\Delta r^2}$) for small, intermediate and large activity numbers $\actNum$ (solid lines) with the corresponding fits (dashed line; \eq{eq:MSD-fit-zottl}) for strongly anchored colloids.
    }
    \label{figsi:MSDfits}
\end{figure}

To enforce no-slip boundary conditions, we prescribe $M_{\vel, \mathrm{T}} = M_{\vel, \mathrm{N}} = -1$ (\eq{eq:mpcd-bc-vel-rule}), as well as including ghost particles to mitigate against artificially lowered viscosities near the surface~\cite{GompperIhle2009Book-MPCD}. 
The unanchored boundary condition corresponds to not altering orientation of colliding particles, \ie $M_{\ori, \mathrm{T}} = M_{\ori, \mathrm{N}} = 1$ (\eq{eq:mpcd-bc-ori-rule}).
To enforce homeotropic anchoring, the orientation of particles is set to match the local surface normal $\norm$ at the point of collision by setting $M_{\ori, \mathrm{T}} = 0$ and $M_{\ori, \mathrm{N}} = 1$ (\eq{eq:mpcd-bc-ori-rule}).
However, the orientation rule presented in \eq{eq:mpcd-bc-ori-rule} only applies \emph{weak} anchoring to the boundary surface~\cite{head2024}. 
To accomplish \emph{strong} anchoring, it is sufficient to apply the orientation collision rule to all particles contained within an MPCD cell that intersects with the boundary surface~\cite{head2024}.

Conservation of linear and angular momentum between colliding particles and the boundary surface allows for the colloidal disk to move in response to the solvent~\cite{head2024}:
For each particle colliding with the boundary surface, the change in linear and angular momentum that occur due to the collision rules is recorded, applying an equal but opposite impulse to the boundary surface.
Resolving the net impulse allows the colloid to move in response to the solvent.
Angular momentum changes of nematogens are allowed because a \emph{virtual length} is prescribed to the point particles~\cite{head2024}. 

\subsubsection{System setup and simulation parameters}\label{si:sec:parameters}
%
Natural simulation units for Andersen-thermostatted MPCD are defined in terms of the cell length $\cellLen = 1$, fluid particle mass $\mass = 1$ and thermal energy $\kbt = 1$. 
The resultant time unit is derived from the natural units to be $\tau = a\sqrt{\mass/\kbt}=1$ and the timestep $\dt$ is given in units of $\tau$. 
Simulations are performed in two dimensions, with an average number of MPCD particles per cell of $\av{\mathcal{N}_\cell}=20$ and timestep size $\dt=0.1\tau$.
The nematic mean field parameter is set to $\mfp=10$ which puts the system firmly in the nematic phase.
The Jeffery parameters governing the tumbling behavior of MPCD particles under the nematic collision operator include a tumbling parameter of $\lambda=2$, which places the simulation in the shear-aligning regime. In addition, a rotational friction coefficient of $\chi_\mathrm{HI}=0.01$ is used, while a hydrodynamic susceptibility of $\chi=0.5$ controls the degree of alignment relaxation relative to $\dt$.
The virtual rod length of the nematogens is set to $\ell_\mathrm{nem}=0.007$.

The system has periodic boundaries, and we vary the two-dimensional simulation box length $\lenSys\in[70, 200]$ linearly with the colloidal radius $\colRad\in[4,35]$.
This choice minimizes computational cost while ensuring the system domain remains sufficiently large to reduce finite-size effects. 
The mass of the colloid is set such that it has the same mass density as the solvent, \ie $M_\mathrm{C}=\pi m \colRad^2 \av{\mathcal{N}_\cell}$. 
The activity $\act$ is varied within the extensile effective turbulence regime~\cite{Kozhukhov2024-ANSig}, $\act \in [0.04,0.27]$, to give a range of active length scales $\lenAct \in [5, 15]$, as measured by the average spacing between nematic defects in bulk turbulence~\cite{Hemingway2016SoftMatter}.
Our choices of $\lenAct$ and $\colRad$ give a set of 10 logarithmically spaced activity numbers $\actNum$ (\eqN{eq:actNum}) in the range $\actNum\in[0.03,2.37]$.

Simulations proceed through a warm-up period of $10^4$ MPCD timesteps, during which the solvent reaches a steady-state topological turbulence while the colloid is pinned in the center of the domain, followed by $3\times 10^5$ timesteps of data collection in which the colloid is allowed to move.
Each activity number $\actNum$ is run for at least 40 individual realizations with random seeds, and results are averaged between these repeats to ensure statistical significance.
For each colloidal radius $\colRad$, a corresponding set of 20 passive nematic simulations are ran. This allows for comparisons between colloids in passive and active nematics.

\begin{figure*}[tb]
    \centering
    \hfill
    \sidesubfloat[]{%
        \includegraphics[width=0.45\linewidth]{Figures/Passive/passive_diffusivity_L.pdf}
    }
    \hfill
    \sidesubfloat[]{%
        \includegraphics[width=0.45\linewidth]{Figures/MSD/msd_fit_eff_diffusivity_vbar.pdf}
    }
    \bcaption{Colloidal diffusion in active nematics exhibits non-monotonic behavior when compared to passive nematics.}{
        \textbf{(a)} 
        Measured Brownian diffusivity $\DBrownian$ from mean-square displacements for colloids of different radii in a passive nematic.
        \textbf{(b)} 
        Measured effective diffusivity $D = \DBrownian + v_0^2 \tau_\mathrm{r} / d$ for different colloids in an active nematic, with activity number $\actNum\simeq\colRad^2/\lenAct^2$.
    }
    \label{figsi:MSD-diffs}
\end{figure*}

\subsection{Mean-Squared Displacement of Colloids}\label{si:sec:msd}
The trajectories of colloidal particles $\pos (t)$ as a function of time $t$ can be quantified by their mean-squared displacement (MSD). 
The MSD is fitted with
\begin{equation} \label{eq:MSD-fit-zottl}
    \av{\Delta r^2}(\dt) = 2d\DBrownian\dt + 2 v_0^2 \tau_\mathrm{r} \dt - 2v_0^2 \tau_\mathrm{r}^2 (1-e^{- \dt/\tau_\mathrm{r}}),
\end{equation}
where $\av{\Delta r^2}(\dt)=\av{\left(\vec{r}(t)-\vec{r}(t+\dt)\right)^2}$, $\dt$ is the lag time, $d$ is the dimensionality of the system, $\DBrownian$ is a passive Brownian diffusivity, $v_0$ is an instantaneous advective speed of the colloidal particle (\fig{figsi:MSDfits}).
The relaxation time $\tau_\mathrm{r}$ describes the persistence time over which the motion de-correlates into a random walk.
As a model for self-propelled particles~\cite{Zottl2016JCP}, \eq{eq:MSD-fit-zottl} describes both short-time propulsive and long-time diffusive limits, with the long-time behavior described by an effective diffusivity $D \equiv \DBrownian + v_0^2 \tau_\mathrm{r} / d$, which combines the passive size-dependent Brownian diffusivity $\DBrownian$ and an active contribution $v_0^2 \tau_\mathrm{r} / d$. 
Brownian diffusivity $D_0$ is extracted from the passive simulations (\fig{figsi:MSD-diffs}a), leaving two fitting parameters $v_0$ and $t_r$, which are extracted using a non-linear least squares algorithm. 
The non-normalized effective diffusivity (\fig{figsi:MSD-diffs}b) exhibits a peak at the same activity number as the normalized curve (\figN{fig:fig2rework}c). 

\begin{figure*}[tb]
    \centering
    \sidesubfloat[]{%
        \includegraphics[width=0.45\linewidth]{Figures/SI_Misc/def_rdf.pdf}
    }
    \sidesubfloat[]{%
        \includegraphics[width=0.45\linewidth]{Figures/SI_Misc/neg_def_dist.pdf}
    }
    \hfill
    \sidesubfloat[]{%
        \includegraphics[width=0.45\linewidth]{Figures/SI_Misc/def_means.pdf}
    }
    \bcaption{Activity restructures the local topological environment in the colloid surroundings. }{
        \textbf{(a)}
        Radial density function $g(r)$ of $-1/2$ defects around the colloid, such that the number density of $-1/2$ defects is $\rho_{-}(r) = g(r) \av{\rho_\mathrm{d}}$, for activity numbers $\actNum\sim\colRad^2/\lenAct^2$.
        \textbf{(b)}
        Probability distributions of the distance to the \emph{nearest} $-1/2$ defect to the colloid across all timesteps for varying activity numbers $\actNum\sim\colRad^2/\lenAct^2$.
        \textbf{(c)}
        Mean distance to the \emph{nearest} $-1/2$ and $+1/2$ defects to the colloid across all timesteps.
        Shaded region represents the measured standard deviation, and the dashed line represents $\actNum=\actCrit$.
    }
    \label{figsi:def-radial-dist}
\end{figure*}

\subsection{Non-anchored colloid}\label{si:sec:non-anchored-colloid}
In contrast to a colloid with homeotropic anchoring, which carries an effective topological charge of $+1$, a colloid with no anchoring exhibits no net topological charge. 
Consequently, the topologically neutral unanchored colloid does not generate companion defects in nematic fluids~\cite{yuan2024chiral}, nor does it display the non-monotonic behavior observed for anchored colloids in active turbulence (\figN{fig:fig2rework}).
To compare with existing literature, we highlight previous work by Lagarde \etal~\cite{Lagarde2020SoftMatter}, who have considered experiments and simulations of unanchored colloidal particles in a bacterial bath.
In their study, the colloidal radius $\colRad$ was fixed, while the bacterial concentration, which acts as a proxy for the activity $\act$.
Similar to the results reported here, they observed plateauing behavior in both the instantaneous colloidal velocity (\figN{fig:fig2rework}a) and the effective diffusion (\fig{figsi:MSD-diffs}b).
However, their decorrelation time $\tau_\mathrm{r}$ remains constant for their simulations and monotonically increasing for experiments.
In contrast, our simulations reveal that $\tau_\mathrm{r}$ plateaus around $\tau_\mathrm{r} \simeq 10$, with minor fluctuations and a slight downward trend for $\actNum\gtrsim\actCrit$ (\figN{fig:fig2rework}b).

\subsection{Companion defects}\label{si:sec:companion-defects}

We posit that the observed changes in colloidal dynamics stem from the evolving structure of the colloidal complex, driven by the interplay between activity-induced bulk defects and colloid-associated companion defects.
The change in the complex is quantified using the radial distribution function (RDF) of $-1/2$ defects around the colloid $g(r) = \av{\rho_\mathrm{n}(r)}/\av{\rho_\mathrm{d}}$, where $\rho_\mathrm{n}(r)$ is the $-1/2$ defect number density found at a distance $r$ from the colloid, and $\av{\rho_\mathrm{d}}$ is the average density of defects, which is non-zero in topological turbulence.
For low activities, the RDF peaks near the colloid (\fig{figsi:def-radial-dist}a), indicative of companion defects, but then becomes lower at a distance from the colloid, indicating a depletion region. 
As activity further increases, the distribution of $-1/2$ defects becomes more homogeneous.

We quantify companion defects themselves by identifying them as the closest $-1/2$ defect to the colloid (\fig{figsi:def-radial-dist}b).
For $-1/2$ companion defects in the passive limit, the distance to companion defects scales with the colloid radius $\colRad$~\cite{Muvsevivc2017-LiquidCrystalColloidsBook}.
However, in the case of topological turbulence, companion defects experience a competition between active and anchoring forces:
The active forces dominate over the anchoring forces, pushing companion defects closer to the colloidal surface with increasing activity:
The mean distance from the colloidal surface to the nearest $-1/2$ and $+1/2$ defects both decrease with increasing activity (\fig{figsi:def-radial-dist}c).
Thus, we conclude companion $-1/2$ defects are found within a distance of $r\lesssim \defDistNeg = 5$.

Defining the near colloid region as $r\leq \defDistNeg$ allows us to quantify the number of companion defects $N_\mathrm{C}$ within this region (\fig{figsi:companion-radial-dist}).
The distribution of $N_\mathrm{C}$ shifts to larger values and broadens as $\actNum$ increases above $\actCrit$ (\fig{figsi:companion-radial-dist}).
The corresponding mean show two clear regimes (\figN{fig:fig3rework}a):
at low activity numbers, the average number of $-1/2$ companion defects lies slightly below one, with a high probability of finding no defects or just a single defect, which matches the qualitative observations at low activity numbers (\figN{fig:fig1A}a.i, \movie{mov:dir-lowA}).
However, at higher activities, the number of defects surrounding the colloid rises steadily, and the structure of the surrounding defects becomes more gas-like.
In stark contrast to the case of low activity numbers, where individual companion defects are identifiable in the RDF (\fig{figsi:def-radial-dist}a), the high density of local defects means that the notion of individual companion defects breaks down at higher activities.
Instead, the probability distribution of the number of $-1/2$ defects within this region has a larger mean and a broader distribution, with peaks occurring at values larger than one (\fig{figsi:companion-radial-dist}).

\begin{figure}[tb]
    \centering
    \includegraphics[width=0.7\linewidth]{Figures/SI_Misc/companion_dist.pdf}
    \bcaption{Distribution of the number of $-1/2$ defects found within $\defDistNeg = 5$ of the colloid surface, $N_\mathrm{C}$, for varying activity number $\actNum\sim\colRad^2/\lenAct^2$.}{
        This is equivalent to the number of potential companion defects to the colloid.
    }
    \label{figsi:companion-radial-dist}
\end{figure}

\subsection{Companion defect ejection}\label{si:sec:ejected}

The observations of only single companion defects in the colloid-companion complex for low activities are at odds with expectations of two companion defects in the limit $\actNum\to0$~\cite{tkalec2013topology}.
 This discrepancy arises from the interaction between the colloid and surrounding activity-induced bulk defects in topological turbulence. 
 The ejection of one colloid-associated companion $-1/2$ defect is facilitated by the presence of nearby activity-induced $+1/2$ defects. 
This process can be qualitatively understood by considering the elastic interactions between the $-1/2$ companion defect, the $+1$ topologically charged colloid, the remaining $-1/2$ defect, and an approaching $+1/2$ defect (\fig{figsi:theory-barrier-height}a). 
For simplicity, we consider the case in which all three defects and the center of the colloid lie along the same line, though the argument qualitatively stands for other configurations. 
Setting the origin on the surface of the colloid closest to the ejected companion defect, we fix the positions of the colloid and remaining companion defect at $-R_C$ and $-2.1 R_C$, respectively, values that are consistent with simulation results. 
The ejected, negative, companion defect is at $r_\mathrm{n}$ and the approaching positive, activity-induced defect is at $r_\mathrm{p}$. 

The elastic energy barrier that the ejected defect must overcome to escape the complex is approximated by their far field expressions. 
This elastic energy barrier is obtained from the total elastic energy,
\begin{equation}\label{eq:companion-def-potential}
    \begin{split}
        U(\vec{r}_\mathrm{n}) = 
        &
        U_{+1, -1/2} + U_{-1/2, -1/2}+U_{+1/2, -1/2}+U_\mathrm{d}(\abs{\vec{r}_\mathrm{n}}),
    \end{split}
\end{equation}
in which $U_\mathrm{d}(r) = -\frac{\pi}{4} K\ln\left(1-\frac{\colRad^2}{r^2}\right)$ comes from the strong anchoring of the colloid that causes an additional short-range repulsion between the companion and the colloid~\cite{Loewe2021NJP}, and 
\begin{equation}
 \label{eq:defect-potential}
      U_{k_1 k_2}(\vec{r}_1,\vec{r}_1) = -2\pi K k_1 k_2 \ln(\abs{\vec{r}_1-\vec{r}_2}),
 \end{equation}
which is the elastic energy between two entities of topological charges $k_1$ and $k_2$ at positions $\vec{r}_1$ and $\vec{r}_2$. 
The first term in \eq{eq:companion-def-potential} represents the interaction between the colloid and the ejected companion defect, whereas the second and third terms represents the interactions between the ejected companion defect and the remaining companion defect, and the approaching $+1/2$ defect, respectively.

When the $+1/2$ defect is far from the colloid ($r_\mathrm{p}=9\colRad$), the energy barrier is large (\fig{figsi:theory-barrier-height}b), which keeps the companion defect bound the colloid complex. 
However, when the $+1/2$ defect approaches the colloid, the potential barrier shrinks significantly, making it possible for the ejected colloid to escape, being effectively snatched by the approaching $+1/2$ defect. 
In fact, as a $+1/2$ defect approaches the colloid, the depth of the potential barrier felt by the $-1/2$ companion decreases monotonically with $r_\mathrm{p}$ (\fig{figsi:theory-barrier-height}c), eventually disappearing entirely if the $+1/2$ defect gets close enough ($r_\mathrm{p}=2.66\colRad$). 
This suggests that a $+1/2$ defect can approach the colloid-companion complex and allow a companion defect to escape from the near-colloid vicinity into the bulk topological turbulence, leaving a single companion defect bound to the colloid.

However, the loss of one companion defect enhances the elastic binding on the remaining companion defect:
Repeating the above analysis for the case where there is only a single companion defect, the potential barrier felt by the remaining companion defect is significantly increased (\fig{figsi:theory-barrier-height}c).
With only a single companion defect, the colloid-companion complex has a non-neutral charge of $+1/2$, which repels any approaching $+1/2$ defects.
This effect is not accounted for in the model (\eq{eq:companion-def-potential}), and so serves as an additional effect beyond the increased energetic barrier, further disfavoring the remaining companion defect from escaping.

\begin{figure*}[tb]
    \centering
    \hfill
    \sidesubfloat[]{%
        \includegraphics[width=0.1\linewidth]{Figures/SI_Misc/Schematic.pdf}
    }
    \hfill
    \sidesubfloat[]{%
        \includegraphics[width=0.38\linewidth]{Figures/SI_Misc/theory_plot.pdf}
    }
    \hfill
    \sidesubfloat[]{%
        \includegraphics[width=0.38\linewidth]{Figures/SI_Misc/barrier_height.pdf}
    }
    \bcaption{Approaching $+1/2$ defects result in one companion $-1/2$ defect being ejected.}{ 
        \textbf{(a)} 
        Schematic of the theoretical model used in \sect{si:sec:ejected}. 
        A $+1/2$ defect approaches the colloid/pair-defect complex of neutral charge from below.
        The $+1/2$ defect is a distance $r_\mathrm{p}$ away from the complex, and the $-1/2$ defect is a distance $r_\mathrm{n}$ away from the complex.
        The other companion defect is at the opposite side of the colloid at $r=-2.1\colRad$.
        \textbf{(b)}
        The potential felt by a $-1/2$ defect $r_\mathrm{n}$ away from a homeotropically anchored colloid surface as a $+1/2$ defect approaches the complex from a distance of $r_\mathrm{p}$.
        It is assumed another $-1/2$ defect is present on the opposite side at $r=-2.1\colRad$. 
        In the main panel there are $N_\mathrm{C}=2$ minus half companion defects.
        \textbf{(c)}
        Measured height of the potential barrier in (b) as a function of the distance to the approaching $+1/2$ defect. 
        The barrier height for both the case of two companion defects ($N_\mathrm{C}=2$) and one companion defect ($N_\mathrm{C}=1$) are shown.
    }
    \label{figsi:theory-barrier-height}
\end{figure*}

\subsection{Quantification of -1/2 defects in colloid reference frame}\label{si:sec:quant-defects}

\begin{figure*}
    \centering 
    \sidesubfloat[]{%
    \includegraphics[width=0.3\linewidth]{Figures/Defect Heatmaps/vel_radial_neg_B01_cbar.pdf}
    }
    \sidesubfloat[]{%
    \includegraphics[width=0.3\linewidth]{Figures/Defect Heatmaps/vel_radial_neg_B02_cbar.pdf}
    }
    \sidesubfloat[]{%
    \includegraphics[width=0.3\linewidth]{Figures/Defect Heatmaps/vel_radial_neg_B03_cbar.pdf}
    }
    
    \sidesubfloat[]{%
    \includegraphics[width=0.3\linewidth]{Figures/Defect Heatmaps/vel_radial_neg_B04_cbar.pdf}
    }
    \sidesubfloat[]{%
    \includegraphics[width=0.3\linewidth]{Figures/Defect Heatmaps/vel_radial_neg_B05_cbar.pdf}
    }
    \sidesubfloat[]{%
    \includegraphics[width=0.3\linewidth]{Figures/Defect Heatmaps/vel_radial_neg_B06_cbar.pdf}
    }

    \sidesubfloat[]{%
    \includegraphics[width=0.3\linewidth]{Figures/Defect Heatmaps/vel_radial_neg_B07_cbar.pdf}
    }
    \sidesubfloat[]{%
    \includegraphics[width=0.3\linewidth]{Figures/Defect Heatmaps/vel_radial_neg_B08_cbar.pdf}
    }
    \sidesubfloat[]{%
    \includegraphics[width=0.3\linewidth]{Figures/Defect Heatmaps/vel_radial_neg_B09_cbar.pdf}
    }
    
    \sidesubfloat[]{%
    \includegraphics[width=0.3\linewidth]{Figures/Defect Heatmaps/vel_radial_neg_B10_cbar.pdf}
    }

    \bcaption{Activity changes the distribution of $-1/2$ defects in the colloid vicinity both angularly and radially.}{ 
        Distribution of $-1/2$ defects in the moving colloid reference frame, where the colloidal velocity $\colVel$ is upwards, for a range of activity numbers $\actNum\simeq \colRad^2/\lenAct^2$.
        The colloidal object is drawn as a blue circle.
    }
    \label{figsi:def-heatmaps}
\end{figure*}

To quantify the position of the remaining $-1/2$ companion defect in the reference frame of the moving colloid, we measure the distribution density function (\fig{figsi:def-heatmaps}) and consider the angle $\theta$ that nearby defects make with respect to the colloidal velocity $\colVel$ (\fig{figsi:def-ang-dist}a)
We refer to $-1/2$ defects in front of the colloid motion as being in the \emph{upper hemisphere} ($|\theta| < \pi / 2$), whereas those that lie behind are in the \emph{lower hemisphere} ($|\theta| > \pi / 2$).
As with the definition of the number of companion defects $N_\mathrm{C}$ (\figN{fig:fig3rework}a; \sect{si:sec:companion-defects}), we consider the range $r < \defDistNeg = 5$ to define the defects that are in the vicinity of the colloid surface.
The degree of symmetry breaking is quantified by considering the angular distribution of nearby $-1/2$ defects (\fig{figsi:def-ang-dist}a).
This reveals two distinct behaviors:
at the lowest activity numbers ($\actNum<\actCrit$), the angular distribution of $-1/2$ defects is heavily biased to the lower hemisphere with a strong preference for defects to reside at the south pole ($|\theta| = \pi$; \fig{figsi:def-ang-dist}a).
As the activity number increases above $\actNum\simeq \actCrit$, the colloid is surrounded by a gas of defects and the lower hemisphere bias begins to decrease, with defects beginning to be more homogeneously spread around the colloid.

This discussion can be more concisely expressed through the difference in the fraction of $-1/2$ defects found in the upper ($0 < |\theta| < \pi/2$) and lower ($\pi/2 < |\theta| < \pi$) hemispheres (\fig{figsi:def-ang-dist}b).
For low activity numbers $\actNum <\actCrit$, there is a strong bias for $-1/2$ defects to be found in the lower hemisphere, with this bias remaining similar as the activity number increases.
However, once $\actNum\simeq\actCrit$, this bias begins to decrease, with the fraction increasing to zero.
the rise continues for $\actNum\gtrsim\actCrit$ past the point where $-1/2$ defects are homogeneous around the colloid.
At the highest activities, while overall homogeneous there is a small bias for defects to lie in the upper hemisphere.

\begin{figure*}
    \hfill
    \sidesubfloat[]{%
    \includegraphics[width=0.45\linewidth]{Figures/SI_Misc/def_angle_dist_lines_neg.pdf}
    }
    \hfill
    \sidesubfloat[]{%
    \includegraphics[width=0.45\linewidth]{Figures/SI_Misc/def_hemisphere_count_2line_neg.pdf}
    }
    \hfill
    \sidesubfloat[]{%
    \includegraphics[width=0.45\linewidth]{Figures/SI_Misc/def_angle_dist_lines_pos_long.pdf}
    }
    \hfill
    \sidesubfloat[]{%
    \includegraphics[width=0.45\linewidth]{Figures/SI_Misc/def_hemisphere_count_2line_pos_long.pdf}
    }
    \bcaption{Angular distributions defects in the reference frame of the colloidal velocity.}{
        \textbf{(a)}
        Distributions of the angle made between companion $-1/2$ defects within $\defDistNeg=5$ of the colloid surface relative to the colloidal instantaneous velocity $\colVel$.
        The dashed line represents $\theta=\pi/2$, the midpoint/ equator of the colloid, and the dotted line represents the mean probability when the number of defects are angularly homogeneous around the colloid.
        \textbf{(b)} 
        The disparity between the number of $-1/2$ defects within $\defDistNeg=5$ of the colloid on top and the bottom of the colloidal midpoint/ equator.
        Mathematically, for the number of defects on top ($N_\uparrow$; $\theta <\pi/2$) and the number of defects on the bottom ($N_\downarrow$; $\theta>\pi/2$), this is $(N_\uparrow - N_\downarrow)/(N_\uparrow + N_\downarrow)$.
        The dashed line represents $\actNum=\actNum^*$, and the dotted line represents 0, where the number of defects above and below the colloidal midpoint are equal.
        \textbf{(c)} 
        Same as (a) but for $+1/2$ defects within $\defDistPos=15$ of the colloid surface.
        \textbf{(d)}
        Same as (a) but for $+1/2$ defects within $\defDistPos=15$ of the colloid surface.
    }
    \label{figsi:def-ang-dist}
\end{figure*}

\subsection{Quantification of +1/2 defects in colloid reference frame}\label{si:sec:quant-defects-pos}

To better understand why the colloidal speed saturates and the diffusivity decreases at high activity numbers, we consider the distribution of $+1/2$ defects in the moving colloidal reference frame, akin to the analysis done in \sect{si:sec:quant-defects}.
The mean distances to the nearest $+1/2$ defects to the colloid (\fig{figsi:def-radial-dist}c) indicate that a \emph{near-colloid} region for approaching $+1/2$ defects can be defined as being within $\defDistPos = 15$ of the colloid surface.
The angular distributions of $+1/2$ defects in this region reveal they are not uniformly distributed around the colloid (\fig{figsi:def-ang-dist}c).
Rather, $+1/2$ defects are preferentially found in the lower hemisphere at low activity numbers but are slightly biased to the upper hemisphere at high activity numbers.
However, more quantitative differences can be found by considering the difference in the fraction of $+1/2$ defects found in the upper and lower hemispheres (\fig{figsi:def-ang-dist}d).
Compared with the $-1/2$ defects, the $+1/2$ defects have a more pronounced bias towards the upper hemisphere, with the transition from a lower to an upper hemisphere bias occurring at a lower activity number than for $-1/2$ defects.

\subsection{Deflection of incoming $+1/2$ defects}\label{si:sec:deflection}

Far from the colloid, an incoming activity-induced $+1/2$ defect sees the colloid-companion complex with a single defect as a net $+1/2$ charge. 
However, the effective repulsion of the complex is weak and is overcome by the active self-propulsion of the defect, leading to a net effective attraction. 
As the defect approaches the colloid, it becomes susceptible to the short-range structure of the dipole potential. 
This furthers the attraction towards the negative charge component of the dipole. 
However, as the self-motile $+1/2$ defect approaches the complex, the repulsive component due to the strong homeotropic anchoring becomes dominant, creating an energy barrier that deflects the incoming $+1/2$ defect and thus shields the companion defect from annihilation.
This section explains in detail the effective potential felt by a $+1/2$ defect as it approaches the colloid-defect complex. 

The effective potential is constructed from both elastic and active forces acting on the colloid.
The approach is similar to the one taken in \sect{si:sec:ejected} to model the elastic potential felt by the $+1/2$ defect approaching the colloid-companion complex. 
However, to allow deflection, the defects and colloid are no longer constrained to fall on a single line. 
The single remaining $-1/2$ companion defect is fixed at the position $\vec r_\mathrm{n}=0.1\colRad \unitvec{i}$ and the colloid at $-\colRad\unitvec{i}$, such that its surface lies at the origin. 
The self-motile $+1/2$ defect is at $\vec{r}_\mathrm{p}$.
The elastic potentials felt by the $+1/2$ defect are those caused by the $+1$ colloid and its $-1/2$ companion defect and are given by \eq{eq:defect-potential}. 
These result in repulsive and attractive potentials, respectively.
For simplicity, we model the self-motility of the $+1/2$ defect as a constant force pointing in the direction of the colloid ($+\unitvec{i}$).
It can thus be modeled as a linear potential, $\vec{F}_\act = - \grad U_\act = \grad (a_\act \vec{r}_\mathrm{p}\cdot \unitvec{i})$, where $a_\act$ quantifies the strength of the $+1/2$ self-propulsion due to the extensile activity $\act$. 

In a similar vein to the construction of \eq{eq:companion-def-potential}, the effective potential felt by the self-propelled $+1/2$ defect is then given by a combination of both the elastic forces and the active forcing
\begin{equation}\label{eq:pos-def-potential}
    \begin{split}
        U_\mathrm{eff}(\pos_\mathrm{p}) = 
        &
        U_{+1, +1/2}(-\colRad\unitvec{i},\pos_\mathrm{p}) + U_{-1/2, +1/2}(0.1 \colRad\unitvec{i}, \pos_\mathrm{p})
        \\
        &
        +U_\mathrm{d}(\abs{\pos_\mathrm{p}}) + a_\act \pos_\mathrm{p} \cdot \unitvec{i} . 
    \end{split}
\end{equation}
The resulting potential experienced by a $+1/2$ defect (\eq{eq:pos-def-potential}) is saddle-like in nature (\figN{fig:fig4rework}b).
As the activity parameter $a_\act$ increases, the saddle point approaches closer to the colloidal surface (\fig{figsi:theory-saddle-dist}).
This persists until $a_\act=0.29$, where the saddle lies atop the $-1/2$ companion defect, at which point there remains no barrier preventing annihilation.

As \eq{eq:pos-def-potential} already includes active motion through the last term, an approaching $+1/2$ defect can be modeled as an overdamped Langevin trajectory 
\begin{equation}
    \deriv{\pos_\mathrm{p}}{t} = - \mu \grad U_\mathrm{eff} + \sqrt{2D_\mathrm{p}}\vec\noise(t),
\end{equation}
where $\mu=D/\kbt$, $D_\mathrm{p}$ is the diffusion coefficient for the model $+1/2$ defect, and $\vec\noise(t)$ is a white noise term that captures the effects of fluctuating hydrodynamics.
A realization of a Langevin trajectory for an approaching $+1/2$ defect is shown as a black line in \figN{fig:fig4rework}b, for $\mu=1$ and $D_\mathrm{p}=0.001$. 

\begin{figure}
    \includegraphics[width=0.7\linewidth]{Figures/Theory/saddle_point_distance.pdf}
    \bcaption{The distance from the colloid surface to the saddle point obtained from \eq{eq:pos-def-potential} for varying activity $a_\act$.}{
        At maximum activity ($a_\act=0.29$), the saddle point lies on the position of the $-1/2$ companion defect, indicated by the blue dot.
    }
    \label{figsi:theory-saddle-dist}
\end{figure}


\clearpage

\bibliography{biblio}




%% file: custom_commands.tex
\usepackage{xspace}

\renewcommand{\vec}[1]{\boldsymbol{#1}}

\DeclareDocumentCommand\corr{ m g }{%
    {C_{#1,#1}%
        \IfNoValueF {#2} { \left(#2\right)}%
    }%
}
\DeclareDocumentCommand\xCorr{ m m g }{%
    {C_{#1,#2}%
        \IfNoValueF {#3} { \left(#3\right)}%
    }%
}
\DeclareDocumentCommand\pcc{ m m g }{%
    {\mathcal{P}_{#1,#2}%
        \IfNoValueF {#3} { \left(#3\right)}%
    }%
}

\newcommand{\dt}{\delta t}

\newcommand{\act}{\alpha}

\newcommand{\actNum}{A}
\newcommand{\actCrit}{\actNum^*}
\newcommand{\Deff}{D}
\newcommand{\DBrownian}{D_0}

\newcommand{\colVel}{\vec{v}_0}
\newcommand{\colSpeed}{v_0}
\newcommand{\colRad}{R_\mathrm{C}}

\DeclareDocumentCommand\length{ g }{%
    {\ell%
        \IfNoValueF {#1} { _{#1} }%
    }%
}
\newcommand{\lenAct}{\length{\act}}
\newcommand{\lenSys}{\length{\mathrm{sys}}}

\newcommand{\fig}[1]{Fig.~\ref{#1}}

\newcommand{\ie}{\textit{i.e.}\xspace}

\newcommand{\tocite}[1]{\textcolor{cyan}{\textsuperscript{*}}}

\newcommand{\bcaption}[2]{\caption{\textbf{#1} #2}}

\newcommand{\afilNbi}{Niels Bohr Institute, University of Copenhagen, Blegdamsvej 17, Copenhagen, Denmark.}
\newcommand{\afilSopa}{School of Physics and Astronomy, University of Edinburgh, Peter Guthrie Tait Road, Edinburgh, EH9 3FD, UK.}
\newcommand{\afilChile}{Facultad de Física, Pontificia Universidad Católica de Chile, Santiago 7820436, Chile.}

%% file: crossref_helper.tex
\usepackage{xr}
\makeatletter

\newcommand*{\addFileDependency}[1]{
\typeout{(#1)}
\@addtofilelist{#1}
\IfFileExists{#1}{}{\typeout{No file #1.}}
}\makeatother

\newcommand*{\myexternaldocument}[1]{%
\externaldocument{#1}%
\addFileDependency{#1.tex}%
\addFileDependency{#1.aux}%
}

%% file: AN-Homeo.bbl
\begin{thebibliography}{48}%
\makeatletter
\providecommand \@ifxundefined [1]{%
 \@ifx{#1\undefined}
}%
\providecommand \@ifnum [1]{%
 \ifnum #1\expandafter \@firstoftwo
 \else \expandafter \@secondoftwo
 \fi
}%
\providecommand \@ifx [1]{%
 \ifx #1\expandafter \@firstoftwo
 \else \expandafter \@secondoftwo
 \fi
}%
\providecommand \natexlab [1]{#1}%
\providecommand \enquote  [1]{``#1''}%
\providecommand \bibnamefont  [1]{#1}%
\providecommand \bibfnamefont [1]{#1}%
\providecommand \citenamefont [1]{#1}%
\providecommand \href@noop [0]{\@secondoftwo}%
\providecommand \href [0]{\begingroup \@sanitize@url \@href}%
\providecommand \@href[1]{\@@startlink{#1}\@@href}%
\providecommand \@@href[1]{\endgroup#1\@@endlink}%
\providecommand \@sanitize@url [0]{\catcode `\\12\catcode `\$12\catcode
  `\&12\catcode `\#12\catcode `\^12\catcode `\_12\catcode `\%12\relax}%
\providecommand \@@startlink[1]{}%
\providecommand \@@endlink[0]{}%
\providecommand \url  [0]{\begingroup\@sanitize@url \@url }%
\providecommand \@url [1]{\endgroup\@href {#1}{\urlprefix }}%
\providecommand \urlprefix  [0]{URL }%
\providecommand \Eprint [0]{\href }%
\providecommand \doibase [0]{https://doi.org/}%
\providecommand \selectlanguage [0]{\@gobble}%
\providecommand \bibinfo  [0]{\@secondoftwo}%
\providecommand \bibfield  [0]{\@secondoftwo}%
\providecommand \translation [1]{[#1]}%
\providecommand \BibitemOpen [0]{}%
\providecommand \bibitemStop [0]{}%
\providecommand \bibitemNoStop [0]{.\EOS\space}%
\providecommand \EOS [0]{\spacefactor3000\relax}%
\providecommand \BibitemShut  [1]{\csname bibitem#1\endcsname}%
\let\auto@bib@innerbib\@empty
\bibitem [{\citenamefont {Marchetti}\ \emph {et~al.}(2013)\citenamefont
  {Marchetti}, \citenamefont {Joanny}, \citenamefont {Ramaswamy}, \citenamefont
  {Liverpool}, \citenamefont {Prost}, \citenamefont {Rao},\ and\ \citenamefont
  {Simha}}]{marchetti2013hydrodynamics}%
  \BibitemOpen
  \bibfield  {author} {\bibinfo {author} {\bibfnamefont {M.~C.}\ \bibnamefont
  {Marchetti}}, \bibinfo {author} {\bibfnamefont {J.-F.}\ \bibnamefont
  {Joanny}}, \bibinfo {author} {\bibfnamefont {S.}~\bibnamefont {Ramaswamy}},
  \bibinfo {author} {\bibfnamefont {T.~B.}\ \bibnamefont {Liverpool}}, \bibinfo
  {author} {\bibfnamefont {J.}~\bibnamefont {Prost}}, \bibinfo {author}
  {\bibfnamefont {M.}~\bibnamefont {Rao}},\ and\ \bibinfo {author}
  {\bibfnamefont {R.~A.}\ \bibnamefont {Simha}},\ }\bibfield  {title} {\bibinfo
  {title} {Hydrodynamics of soft active matter},\ }\href@noop {} {\bibfield
  {journal} {\bibinfo  {journal} {Reviews of Modern Physics}\ }\textbf
  {\bibinfo {volume} {85}},\ \bibinfo {pages} {1143} (\bibinfo {year}
  {2013})}\BibitemShut {NoStop}%
\bibitem [{\citenamefont {Sanchez}\ \emph {et~al.}(2012)\citenamefont
  {Sanchez}, \citenamefont {Chen}, \citenamefont {DeCamp}, \citenamefont
  {Heymann},\ and\ \citenamefont {Dogic}}]{sanchez2012spontaneous}%
  \BibitemOpen
  \bibfield  {author} {\bibinfo {author} {\bibfnamefont {T.}~\bibnamefont
  {Sanchez}}, \bibinfo {author} {\bibfnamefont {D.~T.}\ \bibnamefont {Chen}},
  \bibinfo {author} {\bibfnamefont {S.~J.}\ \bibnamefont {DeCamp}}, \bibinfo
  {author} {\bibfnamefont {M.}~\bibnamefont {Heymann}},\ and\ \bibinfo {author}
  {\bibfnamefont {Z.}~\bibnamefont {Dogic}},\ }\bibfield  {title} {\bibinfo
  {title} {Spontaneous motion in hierarchically assembled active matter},\
  }\href@noop {} {\bibfield  {journal} {\bibinfo  {journal} {Nature}\ }\textbf
  {\bibinfo {volume} {491}},\ \bibinfo {pages} {431} (\bibinfo {year}
  {2012})}\BibitemShut {NoStop}%
\bibitem [{\citenamefont {Zhang}\ \emph {et~al.}(2018)\citenamefont {Zhang},
  \citenamefont {Kumar}, \citenamefont {Ross}, \citenamefont {Gardel},\ and\
  \citenamefont {De~Pablo}}]{zhang2018interplay}%
  \BibitemOpen
  \bibfield  {author} {\bibinfo {author} {\bibfnamefont {R.}~\bibnamefont
  {Zhang}}, \bibinfo {author} {\bibfnamefont {N.}~\bibnamefont {Kumar}},
  \bibinfo {author} {\bibfnamefont {J.~L.}\ \bibnamefont {Ross}}, \bibinfo
  {author} {\bibfnamefont {M.~L.}\ \bibnamefont {Gardel}},\ and\ \bibinfo
  {author} {\bibfnamefont {J.~J.}\ \bibnamefont {De~Pablo}},\ }\bibfield
  {title} {\bibinfo {title} {Interplay of structure, elasticity, and dynamics
  in actin-based nematic materials},\ }\href@noop {} {\bibfield  {journal}
  {\bibinfo  {journal} {Proceedings of the National Academy of Sciences}\
  }\textbf {\bibinfo {volume} {115}},\ \bibinfo {pages} {E124} (\bibinfo {year}
  {2018})}\BibitemShut {NoStop}%
\bibitem [{\citenamefont {Maroudas-Sacks}\ \emph {et~al.}(2021)\citenamefont
  {Maroudas-Sacks}, \citenamefont {Garion}, \citenamefont {Shani-Zerbib},
  \citenamefont {Livshits}, \citenamefont {Braun},\ and\ \citenamefont
  {Keren}}]{maroudas2021topological}%
  \BibitemOpen
  \bibfield  {author} {\bibinfo {author} {\bibfnamefont {Y.}~\bibnamefont
  {Maroudas-Sacks}}, \bibinfo {author} {\bibfnamefont {L.}~\bibnamefont
  {Garion}}, \bibinfo {author} {\bibfnamefont {L.}~\bibnamefont
  {Shani-Zerbib}}, \bibinfo {author} {\bibfnamefont {A.}~\bibnamefont
  {Livshits}}, \bibinfo {author} {\bibfnamefont {E.}~\bibnamefont {Braun}},\
  and\ \bibinfo {author} {\bibfnamefont {K.}~\bibnamefont {Keren}},\ }\bibfield
   {title} {\bibinfo {title} {Topological defects in the nematic order of actin
  fibres as organization centres of hydra morphogenesis},\ }\href@noop {}
  {\bibfield  {journal} {\bibinfo  {journal} {Nature Physics}\ }\textbf
  {\bibinfo {volume} {17}},\ \bibinfo {pages} {251} (\bibinfo {year}
  {2021})}\BibitemShut {NoStop}%
\bibitem [{\citenamefont {Dell’Arciprete}\ \emph {et~al.}(2018)\citenamefont
  {Dell’Arciprete}, \citenamefont {Blow}, \citenamefont {Brown},
  \citenamefont {Farrell}, \citenamefont {Lintuvuori}, \citenamefont {McVey},
  \citenamefont {Marenduzzo},\ and\ \citenamefont {Poon}}]{dell2018growing}%
  \BibitemOpen
  \bibfield  {author} {\bibinfo {author} {\bibfnamefont {D.}~\bibnamefont
  {Dell’Arciprete}}, \bibinfo {author} {\bibfnamefont {M.~L.}\ \bibnamefont
  {Blow}}, \bibinfo {author} {\bibfnamefont {A.~T.}\ \bibnamefont {Brown}},
  \bibinfo {author} {\bibfnamefont {F.~D.}\ \bibnamefont {Farrell}}, \bibinfo
  {author} {\bibfnamefont {J.~S.}\ \bibnamefont {Lintuvuori}}, \bibinfo
  {author} {\bibfnamefont {A.~F.}\ \bibnamefont {McVey}}, \bibinfo {author}
  {\bibfnamefont {D.}~\bibnamefont {Marenduzzo}},\ and\ \bibinfo {author}
  {\bibfnamefont {W.~C.}\ \bibnamefont {Poon}},\ }\bibfield  {title} {\bibinfo
  {title} {A growing bacterial colony in two dimensions as an active nematic},\
  }\href@noop {} {\bibfield  {journal} {\bibinfo  {journal} {Nature
  Communications}\ }\textbf {\bibinfo {volume} {9}},\ \bibinfo {pages} {4190}
  (\bibinfo {year} {2018})}\BibitemShut {NoStop}%
\bibitem [{\citenamefont {van~den Berg}\ \emph {et~al.}(2024)\citenamefont
  {van~den Berg}, \citenamefont {Thijssen}, \citenamefont {Nguyen},
  \citenamefont {Sarlet}, \citenamefont {Cordero}, \citenamefont {V{\'a}zquez},
  \citenamefont {Mitarai}, \citenamefont {Doostmohammadi},\ and\ \citenamefont
  {Jauffred}}]{van2024emergent}%
  \BibitemOpen
  \bibfield  {author} {\bibinfo {author} {\bibfnamefont {N.}~\bibnamefont
  {van~den Berg}}, \bibinfo {author} {\bibfnamefont {K.}~\bibnamefont
  {Thijssen}}, \bibinfo {author} {\bibfnamefont {T.~T.}\ \bibnamefont
  {Nguyen}}, \bibinfo {author} {\bibfnamefont {A.}~\bibnamefont {Sarlet}},
  \bibinfo {author} {\bibfnamefont {M.}~\bibnamefont {Cordero}}, \bibinfo
  {author} {\bibfnamefont {A.~G.}\ \bibnamefont {V{\'a}zquez}}, \bibinfo
  {author} {\bibfnamefont {N.}~\bibnamefont {Mitarai}}, \bibinfo {author}
  {\bibfnamefont {A.}~\bibnamefont {Doostmohammadi}},\ and\ \bibinfo {author}
  {\bibfnamefont {L.}~\bibnamefont {Jauffred}},\ }\bibfield  {title} {\bibinfo
  {title} {Emergent collective alignment gives competitive advantage to longer
  cells during range expansion},\ }\href@noop {} {\bibfield  {journal}
  {\bibinfo  {journal} {bioRxiv}\ ,\ \bibinfo {pages} {2024}} (\bibinfo {year}
  {2024})}\BibitemShut {NoStop}%
\bibitem [{\citenamefont {Duclos}\ \emph {et~al.}(2014)\citenamefont {Duclos},
  \citenamefont {Garcia}, \citenamefont {Yevick},\ and\ \citenamefont
  {Silberzan}}]{Duclos2014}%
  \BibitemOpen
  \bibfield  {author} {\bibinfo {author} {\bibfnamefont {G.}~\bibnamefont
  {Duclos}}, \bibinfo {author} {\bibfnamefont {S.}~\bibnamefont {Garcia}},
  \bibinfo {author} {\bibfnamefont {H.~G.}\ \bibnamefont {Yevick}},\ and\
  \bibinfo {author} {\bibfnamefont {P.}~\bibnamefont {Silberzan}},\ }\bibfield
  {title} {\bibinfo {title} {Perfect nematic order in confined monolayers of
  spindle-shaped cells},\ }\href@noop {} {\bibfield  {journal} {\bibinfo
  {journal} {Soft Matter}\ }\textbf {\bibinfo {volume} {10}},\ \bibinfo {pages}
  {2346} (\bibinfo {year} {2014})}\BibitemShut {NoStop}%
\bibitem [{\citenamefont {Copenhagen}\ \emph {et~al.}(2021)\citenamefont
  {Copenhagen}, \citenamefont {Alert}, \citenamefont {Wingreen},\ and\
  \citenamefont {Shaevitz}}]{copenhagen2021topological}%
  \BibitemOpen
  \bibfield  {author} {\bibinfo {author} {\bibfnamefont {K.}~\bibnamefont
  {Copenhagen}}, \bibinfo {author} {\bibfnamefont {R.}~\bibnamefont {Alert}},
  \bibinfo {author} {\bibfnamefont {N.~S.}\ \bibnamefont {Wingreen}},\ and\
  \bibinfo {author} {\bibfnamefont {J.~W.}\ \bibnamefont {Shaevitz}},\
  }\bibfield  {title} {\bibinfo {title} {Topological defects promote layer
  formation in {M}yxococcus xanthus colonies},\ }\href@noop {} {\bibfield
  {journal} {\bibinfo  {journal} {Nature Physics}\ }\textbf {\bibinfo {volume}
  {17}},\ \bibinfo {pages} {211} (\bibinfo {year} {2021})}\BibitemShut
  {NoStop}%
\bibitem [{\citenamefont {Duclos}\ \emph {et~al.}(2017)\citenamefont {Duclos},
  \citenamefont {Erlenk{\"a}mper}, \citenamefont {Joanny},\ and\ \citenamefont
  {Silberzan}}]{duclos2017topological}%
  \BibitemOpen
  \bibfield  {author} {\bibinfo {author} {\bibfnamefont {G.}~\bibnamefont
  {Duclos}}, \bibinfo {author} {\bibfnamefont {C.}~\bibnamefont
  {Erlenk{\"a}mper}}, \bibinfo {author} {\bibfnamefont {J.-F.}\ \bibnamefont
  {Joanny}},\ and\ \bibinfo {author} {\bibfnamefont {P.}~\bibnamefont
  {Silberzan}},\ }\bibfield  {title} {\bibinfo {title} {Topological defects in
  confined populations of spindle-shaped cells},\ }\href@noop {} {\bibfield
  {journal} {\bibinfo  {journal} {Nature Physics}\ }\textbf {\bibinfo {volume}
  {13}},\ \bibinfo {pages} {58} (\bibinfo {year} {2017})}\BibitemShut {NoStop}%
\bibitem [{\citenamefont {Ruider}\ \emph {et~al.}(2024)\citenamefont {Ruider},
  \citenamefont {Thijssen}, \citenamefont {Vannier}, \citenamefont {Paloschi},
  \citenamefont {Sciortino}, \citenamefont {Doostmohammadi},\ and\
  \citenamefont {Bausch}}]{ruider2024topological}%
  \BibitemOpen
  \bibfield  {author} {\bibinfo {author} {\bibfnamefont {I.}~\bibnamefont
  {Ruider}}, \bibinfo {author} {\bibfnamefont {K.}~\bibnamefont {Thijssen}},
  \bibinfo {author} {\bibfnamefont {D.~R.}\ \bibnamefont {Vannier}}, \bibinfo
  {author} {\bibfnamefont {V.}~\bibnamefont {Paloschi}}, \bibinfo {author}
  {\bibfnamefont {A.}~\bibnamefont {Sciortino}}, \bibinfo {author}
  {\bibfnamefont {A.}~\bibnamefont {Doostmohammadi}},\ and\ \bibinfo {author}
  {\bibfnamefont {A.}~\bibnamefont {Bausch}},\ }\bibfield  {title} {\bibinfo
  {title} {Topological excitations govern ordering kinetics in endothelial cell
  layers},\ }\href@noop {} {\bibfield  {journal} {\bibinfo  {journal}
  {bioRxiv}\ ,\ \bibinfo {pages} {2024}} (\bibinfo {year} {2024})}\BibitemShut
  {NoStop}%
\bibitem [{\citenamefont {Senyuk}\ \emph {et~al.}(2012)\citenamefont {Senyuk},
  \citenamefont {Evans}, \citenamefont {Ackerman}, \citenamefont {Lee},
  \citenamefont {Manna}, \citenamefont {Vigderman}, \citenamefont {Zubarev},
  \citenamefont {van~de Lagemaat},\ and\ \citenamefont
  {Smalyukh}}]{senyuk2012shape}%
  \BibitemOpen
  \bibfield  {author} {\bibinfo {author} {\bibfnamefont {B.}~\bibnamefont
  {Senyuk}}, \bibinfo {author} {\bibfnamefont {J.~S.}\ \bibnamefont {Evans}},
  \bibinfo {author} {\bibfnamefont {P.~J.}\ \bibnamefont {Ackerman}}, \bibinfo
  {author} {\bibfnamefont {T.}~\bibnamefont {Lee}}, \bibinfo {author}
  {\bibfnamefont {P.}~\bibnamefont {Manna}}, \bibinfo {author} {\bibfnamefont
  {L.}~\bibnamefont {Vigderman}}, \bibinfo {author} {\bibfnamefont {E.~R.}\
  \bibnamefont {Zubarev}}, \bibinfo {author} {\bibfnamefont {J.}~\bibnamefont
  {van~de Lagemaat}},\ and\ \bibinfo {author} {\bibfnamefont {I.~I.}\
  \bibnamefont {Smalyukh}},\ }\bibfield  {title} {\bibinfo {title}
  {Shape-dependent oriented trapping and scaffolding of plasmonic nanoparticles
  by topological defects for self-assembly of colloidal dimers in liquid
  crystals},\ }\href@noop {} {\bibfield  {journal} {\bibinfo  {journal} {Nano
  Letters}\ }\textbf {\bibinfo {volume} {12}},\ \bibinfo {pages} {955}
  (\bibinfo {year} {2012})}\BibitemShut {NoStop}%
\bibitem [{\citenamefont {Ohzono}\ and\ \citenamefont
  {Fukuda}(2012)}]{ohzono2012zigzag}%
  \BibitemOpen
  \bibfield  {author} {\bibinfo {author} {\bibfnamefont {T.}~\bibnamefont
  {Ohzono}}\ and\ \bibinfo {author} {\bibfnamefont {J.-i.}\ \bibnamefont
  {Fukuda}},\ }\bibfield  {title} {\bibinfo {title} {Zigzag line defects and
  manipulation of colloids in a nematic liquid crystal in microwrinkle
  grooves},\ }\href@noop {} {\bibfield  {journal} {\bibinfo  {journal} {Nature
  Communications}\ }\textbf {\bibinfo {volume} {3}},\ \bibinfo {pages} {701}
  (\bibinfo {year} {2012})}\BibitemShut {NoStop}%
\bibitem [{\citenamefont {Kosterlitz}(2016)}]{kosterlitz2016topological}%
  \BibitemOpen
  \bibfield  {author} {\bibinfo {author} {\bibfnamefont {J.~M.}\ \bibnamefont
  {Kosterlitz}},\ }\bibfield  {title} {\bibinfo {title} {Topological defects
  and phase transitions},\ }\href@noop {} {\bibfield  {journal} {\bibinfo
  {journal} {Nobel Physics Lecture}\ } (\bibinfo {year} {2016})}\BibitemShut
  {NoStop}%
\bibitem [{\citenamefont {Bowick}\ \emph {et~al.}(2022)\citenamefont {Bowick},
  \citenamefont {Fakhri}, \citenamefont {Marchetti},\ and\ \citenamefont
  {Ramaswamy}}]{bowick2022symmetry}%
  \BibitemOpen
  \bibfield  {author} {\bibinfo {author} {\bibfnamefont {M.~J.}\ \bibnamefont
  {Bowick}}, \bibinfo {author} {\bibfnamefont {N.}~\bibnamefont {Fakhri}},
  \bibinfo {author} {\bibfnamefont {M.~C.}\ \bibnamefont {Marchetti}},\ and\
  \bibinfo {author} {\bibfnamefont {S.}~\bibnamefont {Ramaswamy}},\ }\bibfield
  {title} {\bibinfo {title} {Symmetry, thermodynamics, and topology in active
  matter},\ }\href@noop {} {\bibfield  {journal} {\bibinfo  {journal} {Physical
  Review X}\ }\textbf {\bibinfo {volume} {12}},\ \bibinfo {pages} {010501}
  (\bibinfo {year} {2022})}\BibitemShut {NoStop}%
\bibitem [{\citenamefont {Arda{\v{s}}eva}\ and\ \citenamefont
  {Doostmohammadi}(2022)}]{ardavseva2022topological}%
  \BibitemOpen
  \bibfield  {author} {\bibinfo {author} {\bibfnamefont {A.}~\bibnamefont
  {Arda{\v{s}}eva}}\ and\ \bibinfo {author} {\bibfnamefont {A.}~\bibnamefont
  {Doostmohammadi}},\ }\bibfield  {title} {\bibinfo {title} {Topological
  defects in biological matter},\ }\href@noop {} {\bibfield  {journal}
  {\bibinfo  {journal} {Nature Reviews Physics}\ }\textbf {\bibinfo {volume}
  {4}},\ \bibinfo {pages} {354} (\bibinfo {year} {2022})}\BibitemShut {NoStop}%
\bibitem [{\citenamefont {Kralj}\ \emph {et~al.}(2021)\citenamefont {Kralj},
  \citenamefont {Kralj},\ and\ \citenamefont {Kralj}}]{Kralj2018}%
  \BibitemOpen
  \bibfield  {author} {\bibinfo {author} {\bibfnamefont {M.}~\bibnamefont
  {Kralj}}, \bibinfo {author} {\bibfnamefont {M.}~\bibnamefont {Kralj}},\ and\
  \bibinfo {author} {\bibfnamefont {S.}~\bibnamefont {Kralj}},\ }\bibfield
  {title} {\bibinfo {title} {Topological defects in nematic liquid crystals:
  Laboratory of fundamental physics},\ }\href@noop {} {\bibfield  {journal}
  {\bibinfo  {journal} {Physica Status Solidi (a)}\ }\textbf {\bibinfo {volume}
  {218}},\ \bibinfo {pages} {2000752} (\bibinfo {year} {2021})}\BibitemShut
  {NoStop}%
\bibitem [{\citenamefont {Saw}\ \emph {et~al.}(2017)\citenamefont {Saw},
  \citenamefont {Doostmohammadi}, \citenamefont {Nier}, \citenamefont
  {Kocgozlu}, \citenamefont {Thampi}, \citenamefont {Toyama}, \citenamefont
  {Marcq}, \citenamefont {Lim}, \citenamefont {Yeomans},\ and\ \citenamefont
  {Ladoux}}]{Saw2017}%
  \BibitemOpen
  \bibfield  {author} {\bibinfo {author} {\bibfnamefont {T.~B.}\ \bibnamefont
  {Saw}}, \bibinfo {author} {\bibfnamefont {A.}~\bibnamefont {Doostmohammadi}},
  \bibinfo {author} {\bibfnamefont {V.}~\bibnamefont {Nier}}, \bibinfo {author}
  {\bibfnamefont {L.}~\bibnamefont {Kocgozlu}}, \bibinfo {author}
  {\bibfnamefont {S.}~\bibnamefont {Thampi}}, \bibinfo {author} {\bibfnamefont
  {Y.}~\bibnamefont {Toyama}}, \bibinfo {author} {\bibfnamefont
  {P.}~\bibnamefont {Marcq}}, \bibinfo {author} {\bibfnamefont {C.~T.}\
  \bibnamefont {Lim}}, \bibinfo {author} {\bibfnamefont {J.~M.}\ \bibnamefont
  {Yeomans}},\ and\ \bibinfo {author} {\bibfnamefont {B.}~\bibnamefont
  {Ladoux}},\ }\bibfield  {title} {\bibinfo {title} {Topological defects in
  epithelia govern cell death and extrusion},\ }\href@noop {} {\bibfield
  {journal} {\bibinfo  {journal} {Nature}\ }\textbf {\bibinfo {volume} {544}},\
  \bibinfo {pages} {212} (\bibinfo {year} {2017})}\BibitemShut {NoStop}%
\bibitem [{\citenamefont {Kawaguchi}\ \emph {et~al.}(2017)\citenamefont
  {Kawaguchi}, \citenamefont {Kageyama},\ and\ \citenamefont
  {Sano}}]{kawaguchi2017topological}%
  \BibitemOpen
  \bibfield  {author} {\bibinfo {author} {\bibfnamefont {K.}~\bibnamefont
  {Kawaguchi}}, \bibinfo {author} {\bibfnamefont {R.}~\bibnamefont
  {Kageyama}},\ and\ \bibinfo {author} {\bibfnamefont {M.}~\bibnamefont
  {Sano}},\ }\bibfield  {title} {\bibinfo {title} {Topological defects control
  collective dynamics in neural progenitor cell cultures},\ }\href@noop {}
  {\bibfield  {journal} {\bibinfo  {journal} {Nature}\ }\textbf {\bibinfo
  {volume} {545}},\ \bibinfo {pages} {327} (\bibinfo {year}
  {2017})}\BibitemShut {NoStop}%
\bibitem [{\citenamefont {Tkalec}\ and\ \citenamefont
  {Mu{\v{s}}evi{\v{c}}}(2013)}]{tkalec2013topology}%
  \BibitemOpen
  \bibfield  {author} {\bibinfo {author} {\bibfnamefont {U.}~\bibnamefont
  {Tkalec}}\ and\ \bibinfo {author} {\bibfnamefont {I.}~\bibnamefont
  {Mu{\v{s}}evi{\v{c}}}},\ }\bibfield  {title} {\bibinfo {title} {Topology of
  nematic liquid crystal colloids confined to two dimensions},\ }\href@noop {}
  {\bibfield  {journal} {\bibinfo  {journal} {Soft Matter}\ }\textbf {\bibinfo
  {volume} {9}},\ \bibinfo {pages} {8140} (\bibinfo {year} {2013})}\BibitemShut
  {NoStop}%
\bibitem [{\citenamefont {Head}\ \emph
  {et~al.}(2024{\natexlab{a}})\citenamefont {Head}, \citenamefont {Fosado},
  \citenamefont {Marenduzzo},\ and\ \citenamefont {Shendruk}}]{head2024}%
  \BibitemOpen
  \bibfield  {author} {\bibinfo {author} {\bibfnamefont {L.~C.}\ \bibnamefont
  {Head}}, \bibinfo {author} {\bibfnamefont {Y.~A.~G.}\ \bibnamefont {Fosado}},
  \bibinfo {author} {\bibfnamefont {D.}~\bibnamefont {Marenduzzo}},\ and\
  \bibinfo {author} {\bibfnamefont {T.~N.}\ \bibnamefont {Shendruk}},\
  }\bibfield  {title} {\bibinfo {title} {Entangled nematic disclinations using
  multi-particle collision dynamics},\ }\href@noop {} {\bibfield  {journal}
  {\bibinfo  {journal} {Soft Matter}\ }\textbf {\bibinfo {volume} {20}},\
  \bibinfo {pages} {7157} (\bibinfo {year} {2024}{\natexlab{a}})}\BibitemShut
  {NoStop}%
\bibitem [{\citenamefont {{\v{C}}opar}\ \emph {et~al.}(2015)\citenamefont
  {{\v{C}}opar}, \citenamefont {Tkalec}, \citenamefont {Mu{\v{s}}evi{\v{c}}},\
  and\ \citenamefont {{\v{Z}}umer}}]{vcopar2015knot}%
  \BibitemOpen
  \bibfield  {author} {\bibinfo {author} {\bibfnamefont {S.}~\bibnamefont
  {{\v{C}}opar}}, \bibinfo {author} {\bibfnamefont {U.}~\bibnamefont {Tkalec}},
  \bibinfo {author} {\bibfnamefont {I.}~\bibnamefont {Mu{\v{s}}evi{\v{c}}}},\
  and\ \bibinfo {author} {\bibfnamefont {S.}~\bibnamefont {{\v{Z}}umer}},\
  }\bibfield  {title} {\bibinfo {title} {Knot theory realizations in nematic
  colloids},\ }\href@noop {} {\bibfield  {journal} {\bibinfo  {journal}
  {Proceedings of the National Academy of Sciences}\ }\textbf {\bibinfo
  {volume} {112}},\ \bibinfo {pages} {1675} (\bibinfo {year}
  {2015})}\BibitemShut {NoStop}%
\bibitem [{\citenamefont {Wamsler}\ \emph {et~al.}(2024)\citenamefont
  {Wamsler}, \citenamefont {Head},\ and\ \citenamefont
  {Shendruk}}]{Head2024-WavyWalls}%
  \BibitemOpen
  \bibfield  {author} {\bibinfo {author} {\bibfnamefont {K.}~\bibnamefont
  {Wamsler}}, \bibinfo {author} {\bibfnamefont {L.~C.}\ \bibnamefont {Head}},\
  and\ \bibinfo {author} {\bibfnamefont {T.~N.}\ \bibnamefont {Shendruk}},\
  }\bibfield  {title} {\bibinfo {title} {Lock-key microfluidics: simulating
  nematic colloid advection along wavy-walled channels},\ }\href@noop {}
  {\bibfield  {journal} {\bibinfo  {journal} {Soft Matter}\ }\textbf {\bibinfo
  {volume} {20}},\ \bibinfo {pages} {3954} (\bibinfo {year}
  {2024})}\BibitemShut {NoStop}%
\bibitem [{\citenamefont {Modin}\ \emph {et~al.}(2023)\citenamefont {Modin},
  \citenamefont {Ash}, \citenamefont {Ishimoto}, \citenamefont {Leheny},
  \citenamefont {Serra},\ and\ \citenamefont {Aharoni}}]{modin2023tunable}%
  \BibitemOpen
  \bibfield  {author} {\bibinfo {author} {\bibfnamefont {A.}~\bibnamefont
  {Modin}}, \bibinfo {author} {\bibfnamefont {B.}~\bibnamefont {Ash}}, \bibinfo
  {author} {\bibfnamefont {K.}~\bibnamefont {Ishimoto}}, \bibinfo {author}
  {\bibfnamefont {R.~L.}\ \bibnamefont {Leheny}}, \bibinfo {author}
  {\bibfnamefont {F.}~\bibnamefont {Serra}},\ and\ \bibinfo {author}
  {\bibfnamefont {H.}~\bibnamefont {Aharoni}},\ }\bibfield  {title} {\bibinfo
  {title} {Tunable three-dimensional architecture of nematic disclination
  lines},\ }\href@noop {} {\bibfield  {journal} {\bibinfo  {journal}
  {Proceedings of the National Academy of Sciences}\ }\textbf {\bibinfo
  {volume} {120}},\ \bibinfo {pages} {e2300833120} (\bibinfo {year}
  {2023})}\BibitemShut {NoStop}%
\bibitem [{\citenamefont {Shankar}\ \emph {et~al.}(2018)\citenamefont
  {Shankar}, \citenamefont {Ramaswamy}, \citenamefont {Marchetti},\ and\
  \citenamefont {Bowick}}]{shankar2018defect}%
  \BibitemOpen
  \bibfield  {author} {\bibinfo {author} {\bibfnamefont {S.}~\bibnamefont
  {Shankar}}, \bibinfo {author} {\bibfnamefont {S.}~\bibnamefont {Ramaswamy}},
  \bibinfo {author} {\bibfnamefont {M.~C.}\ \bibnamefont {Marchetti}},\ and\
  \bibinfo {author} {\bibfnamefont {M.~J.}\ \bibnamefont {Bowick}},\ }\bibfield
   {title} {\bibinfo {title} {Defect unbinding in active nematics},\
  }\href@noop {} {\bibfield  {journal} {\bibinfo  {journal} {Physical Review
  Letters}\ }\textbf {\bibinfo {volume} {121}},\ \bibinfo {pages} {108002}
  (\bibinfo {year} {2018})}\BibitemShut {NoStop}%
\bibitem [{\citenamefont {Thampi}\ \emph {et~al.}(2013)\citenamefont {Thampi},
  \citenamefont {Golestanian},\ and\ \citenamefont
  {Yeomans}}]{thampi2013velocity}%
  \BibitemOpen
  \bibfield  {author} {\bibinfo {author} {\bibfnamefont {S.~P.}\ \bibnamefont
  {Thampi}}, \bibinfo {author} {\bibfnamefont {R.}~\bibnamefont
  {Golestanian}},\ and\ \bibinfo {author} {\bibfnamefont {J.~M.}\ \bibnamefont
  {Yeomans}},\ }\bibfield  {title} {\bibinfo {title} {Velocity correlations in
  an active nematic},\ }\href@noop {} {\bibfield  {journal} {\bibinfo
  {journal} {Physical Review Letters}\ }\textbf {\bibinfo {volume} {111}},\
  \bibinfo {pages} {118101} (\bibinfo {year} {2013})}\BibitemShut {NoStop}%
\bibitem [{\citenamefont {Doostmohammadi}\ \emph {et~al.}(2018)\citenamefont
  {Doostmohammadi}, \citenamefont {Ign{\'e}s-Mullol}, \citenamefont {Yeomans},\
  and\ \citenamefont {Sagu{\'e}s}}]{doostmohammadi2018active}%
  \BibitemOpen
  \bibfield  {author} {\bibinfo {author} {\bibfnamefont {A.}~\bibnamefont
  {Doostmohammadi}}, \bibinfo {author} {\bibfnamefont {J.}~\bibnamefont
  {Ign{\'e}s-Mullol}}, \bibinfo {author} {\bibfnamefont {J.~M.}\ \bibnamefont
  {Yeomans}},\ and\ \bibinfo {author} {\bibfnamefont {F.}~\bibnamefont
  {Sagu{\'e}s}},\ }\bibfield  {title} {\bibinfo {title} {Active nematics},\
  }\href@noop {} {\bibfield  {journal} {\bibinfo  {journal} {Nature
  Communications}\ }\textbf {\bibinfo {volume} {9}},\ \bibinfo {pages} {3246}
  (\bibinfo {year} {2018})}\BibitemShut {NoStop}%
\bibitem [{\citenamefont {Giomi}\ \emph
  {et~al.}(2014{\natexlab{a}})\citenamefont {Giomi}, \citenamefont {Bowick},
  \citenamefont {Mishra}, \citenamefont {Sknepnek},\ and\ \citenamefont
  {Cristina~Marchetti}}]{giomi2014defect}%
  \BibitemOpen
  \bibfield  {author} {\bibinfo {author} {\bibfnamefont {L.}~\bibnamefont
  {Giomi}}, \bibinfo {author} {\bibfnamefont {M.~J.}\ \bibnamefont {Bowick}},
  \bibinfo {author} {\bibfnamefont {P.}~\bibnamefont {Mishra}}, \bibinfo
  {author} {\bibfnamefont {R.}~\bibnamefont {Sknepnek}},\ and\ \bibinfo
  {author} {\bibfnamefont {M.}~\bibnamefont {Cristina~Marchetti}},\ }\bibfield
  {title} {\bibinfo {title} {Defect dynamics in active nematics},\ }\href@noop
  {} {\bibfield  {journal} {\bibinfo  {journal} {Philosophical Transactions of
  the Royal Society A: Mathematical, Physical and Engineering Sciences}\
  }\textbf {\bibinfo {volume} {372}},\ \bibinfo {pages} {20130365} (\bibinfo
  {year} {2014}{\natexlab{a}})}\BibitemShut {NoStop}%
\bibitem [{\citenamefont {Shankar}\ and\ \citenamefont
  {Marchetti}(2019)}]{shankar2019hydrodynamics}%
  \BibitemOpen
  \bibfield  {author} {\bibinfo {author} {\bibfnamefont {S.}~\bibnamefont
  {Shankar}}\ and\ \bibinfo {author} {\bibfnamefont {M.~C.}\ \bibnamefont
  {Marchetti}},\ }\bibfield  {title} {\bibinfo {title} {Hydrodynamics of active
  defects: From order to chaos to defect ordering},\ }\href@noop {} {\bibfield
  {journal} {\bibinfo  {journal} {Physical Review X}\ }\textbf {\bibinfo
  {volume} {9}},\ \bibinfo {pages} {041047} (\bibinfo {year}
  {2019})}\BibitemShut {NoStop}%
\bibitem [{\citenamefont {Alert}\ \emph {et~al.}(2022)\citenamefont {Alert},
  \citenamefont {Casademunt},\ and\ \citenamefont
  {Joanny}}]{Alert2022AnnRevCondMat}%
  \BibitemOpen
  \bibfield  {author} {\bibinfo {author} {\bibfnamefont {R.}~\bibnamefont
  {Alert}}, \bibinfo {author} {\bibfnamefont {J.}~\bibnamefont {Casademunt}},\
  and\ \bibinfo {author} {\bibfnamefont {J.-F.}\ \bibnamefont {Joanny}},\
  }\bibfield  {title} {\bibinfo {title} {Active turbulence},\ }\href@noop {}
  {\bibfield  {journal} {\bibinfo  {journal} {Annual Review of Condensed Matter
  Physics}\ }\textbf {\bibinfo {volume} {13}},\ \bibinfo {pages} {143}
  (\bibinfo {year} {2022})}\BibitemShut {NoStop}%
\bibitem [{\citenamefont {Ellis}\ \emph {et~al.}(2018)\citenamefont {Ellis},
  \citenamefont {Pearce}, \citenamefont {Chang}, \citenamefont {Goldsztein},
  \citenamefont {Giomi},\ and\ \citenamefont
  {Fernandez-Nieves}}]{ellis2018curvature}%
  \BibitemOpen
  \bibfield  {author} {\bibinfo {author} {\bibfnamefont {P.~W.}\ \bibnamefont
  {Ellis}}, \bibinfo {author} {\bibfnamefont {D.~J.}\ \bibnamefont {Pearce}},
  \bibinfo {author} {\bibfnamefont {Y.-W.}\ \bibnamefont {Chang}}, \bibinfo
  {author} {\bibfnamefont {G.}~\bibnamefont {Goldsztein}}, \bibinfo {author}
  {\bibfnamefont {L.}~\bibnamefont {Giomi}},\ and\ \bibinfo {author}
  {\bibfnamefont {A.}~\bibnamefont {Fernandez-Nieves}},\ }\bibfield  {title}
  {\bibinfo {title} {Curvature-induced defect unbinding and dynamics in active
  nematic toroids},\ }\href@noop {} {\bibfield  {journal} {\bibinfo  {journal}
  {Nature Physics}\ }\textbf {\bibinfo {volume} {14}},\ \bibinfo {pages} {85}
  (\bibinfo {year} {2018})}\BibitemShut {NoStop}%
\bibitem [{\citenamefont {Hardo{\"u}in}\ \emph {et~al.}(2022)\citenamefont
  {Hardo{\"u}in}, \citenamefont {Dor{\'e}}, \citenamefont {Laurent},
  \citenamefont {Lopez-Leon}, \citenamefont {Ign{\'e}s-Mullol},\ and\
  \citenamefont {Sagu{\'e}s}}]{hardouin2022active}%
  \BibitemOpen
  \bibfield  {author} {\bibinfo {author} {\bibfnamefont {J.}~\bibnamefont
  {Hardo{\"u}in}}, \bibinfo {author} {\bibfnamefont {C.}~\bibnamefont
  {Dor{\'e}}}, \bibinfo {author} {\bibfnamefont {J.}~\bibnamefont {Laurent}},
  \bibinfo {author} {\bibfnamefont {T.}~\bibnamefont {Lopez-Leon}}, \bibinfo
  {author} {\bibfnamefont {J.}~\bibnamefont {Ign{\'e}s-Mullol}},\ and\ \bibinfo
  {author} {\bibfnamefont {F.}~\bibnamefont {Sagu{\'e}s}},\ }\bibfield  {title}
  {\bibinfo {title} {Active boundary layers in confined active nematics},\
  }\href@noop {} {\bibfield  {journal} {\bibinfo  {journal} {Nature
  Communications}\ }\textbf {\bibinfo {volume} {13}},\ \bibinfo {pages} {6675}
  (\bibinfo {year} {2022})}\BibitemShut {NoStop}%
\bibitem [{\citenamefont {Guillamat}\ \emph {et~al.}(2018)\citenamefont
  {Guillamat}, \citenamefont {Kos}, \citenamefont {Hardo{\"u}in}, \citenamefont
  {Ign{\'e}s-Mullol}, \citenamefont {Ravnik},\ and\ \citenamefont
  {Sagu{\'e}s}}]{guillamat2018active}%
  \BibitemOpen
  \bibfield  {author} {\bibinfo {author} {\bibfnamefont {P.}~\bibnamefont
  {Guillamat}}, \bibinfo {author} {\bibfnamefont {{\v{Z}}.}~\bibnamefont
  {Kos}}, \bibinfo {author} {\bibfnamefont {J.}~\bibnamefont {Hardo{\"u}in}},
  \bibinfo {author} {\bibfnamefont {J.}~\bibnamefont {Ign{\'e}s-Mullol}},
  \bibinfo {author} {\bibfnamefont {M.}~\bibnamefont {Ravnik}},\ and\ \bibinfo
  {author} {\bibfnamefont {F.}~\bibnamefont {Sagu{\'e}s}},\ }\bibfield  {title}
  {\bibinfo {title} {Active nematic emulsions},\ }\href@noop {} {\bibfield
  {journal} {\bibinfo  {journal} {Science Advances}\ }\textbf {\bibinfo
  {volume} {4}},\ \bibinfo {pages} {eaao1470} (\bibinfo {year}
  {2018})}\BibitemShut {NoStop}%
\bibitem [{\citenamefont {Negro}\ \emph {et~al.}(2025)\citenamefont {Negro},
  \citenamefont {Head}, \citenamefont {Carenza}, \citenamefont {Shendruk},
  \citenamefont {Marenduzzo}, \citenamefont {Gonnella},\ and\ \citenamefont
  {Tiribocchi}}]{negro2025}%
  \BibitemOpen
  \bibfield  {author} {\bibinfo {author} {\bibfnamefont {G.}~\bibnamefont
  {Negro}}, \bibinfo {author} {\bibfnamefont {L.~C.}\ \bibnamefont {Head}},
  \bibinfo {author} {\bibfnamefont {L.~N.}\ \bibnamefont {Carenza}}, \bibinfo
  {author} {\bibfnamefont {T.~N.}\ \bibnamefont {Shendruk}}, \bibinfo {author}
  {\bibfnamefont {D.}~\bibnamefont {Marenduzzo}}, \bibinfo {author}
  {\bibfnamefont {G.}~\bibnamefont {Gonnella}},\ and\ \bibinfo {author}
  {\bibfnamefont {A.}~\bibnamefont {Tiribocchi}},\ }\bibfield  {title}
  {\bibinfo {title} {Topology controls flow patterns in active double
  emulsions},\ }\href@noop {} {\bibfield  {journal} {\bibinfo  {journal}
  {Nature Communications}\ }\textbf {\bibinfo {volume} {16}},\ \bibinfo {pages}
  {1} (\bibinfo {year} {2025})}\BibitemShut {NoStop}%
\bibitem [{\citenamefont {Valeriani}\ \emph {et~al.}(2011)\citenamefont
  {Valeriani}, \citenamefont {Li}, \citenamefont {Novosel}, \citenamefont
  {Arlt},\ and\ \citenamefont {Marenduzzo}}]{valeriani2011colloids}%
  \BibitemOpen
  \bibfield  {author} {\bibinfo {author} {\bibfnamefont {C.}~\bibnamefont
  {Valeriani}}, \bibinfo {author} {\bibfnamefont {M.}~\bibnamefont {Li}},
  \bibinfo {author} {\bibfnamefont {J.}~\bibnamefont {Novosel}}, \bibinfo
  {author} {\bibfnamefont {J.}~\bibnamefont {Arlt}},\ and\ \bibinfo {author}
  {\bibfnamefont {D.}~\bibnamefont {Marenduzzo}},\ }\bibfield  {title}
  {\bibinfo {title} {Colloids in a bacterial bath: simulations and
  experiments},\ }\href@noop {} {\bibfield  {journal} {\bibinfo  {journal}
  {Soft Matter}\ }\textbf {\bibinfo {volume} {7}},\ \bibinfo {pages} {5228}
  (\bibinfo {year} {2011})}\BibitemShut {NoStop}%
\bibitem [{\citenamefont {Lagarde}\ \emph {et~al.}(2020)\citenamefont
  {Lagarde}, \citenamefont {Dag{\`e}s}, \citenamefont {Nemoto}, \citenamefont
  {D{\'e}mery}, \citenamefont {Bartolo},\ and\ \citenamefont
  {Gibaud}}]{lagarde2020colloidal}%
  \BibitemOpen
  \bibfield  {author} {\bibinfo {author} {\bibfnamefont {A.}~\bibnamefont
  {Lagarde}}, \bibinfo {author} {\bibfnamefont {N.}~\bibnamefont {Dag{\`e}s}},
  \bibinfo {author} {\bibfnamefont {T.}~\bibnamefont {Nemoto}}, \bibinfo
  {author} {\bibfnamefont {V.}~\bibnamefont {D{\'e}mery}}, \bibinfo {author}
  {\bibfnamefont {D.}~\bibnamefont {Bartolo}},\ and\ \bibinfo {author}
  {\bibfnamefont {T.}~\bibnamefont {Gibaud}},\ }\bibfield  {title} {\bibinfo
  {title} {Colloidal transport in bacteria suspensions: from bacteria collision
  to anomalous and enhanced diffusion},\ }\href@noop {} {\bibfield  {journal}
  {\bibinfo  {journal} {Soft Matter}\ }\textbf {\bibinfo {volume} {16}},\
  \bibinfo {pages} {7503} (\bibinfo {year} {2020})}\BibitemShut {NoStop}%
\bibitem [{\citenamefont {Loewe}\ and\ \citenamefont
  {Shendruk}(2022)}]{Loewe2021NJP}%
  \BibitemOpen
  \bibfield  {author} {\bibinfo {author} {\bibfnamefont {B.}~\bibnamefont
  {Loewe}}\ and\ \bibinfo {author} {\bibfnamefont {T.~N.}\ \bibnamefont
  {Shendruk}},\ }\bibfield  {title} {\bibinfo {title} {Passive {J}anus
  particles are self-propelled in active nematics},\ }\href@noop {} {\bibfield
  {journal} {\bibinfo  {journal} {New Journal of Physics}\ }\textbf {\bibinfo
  {volume} {24}},\ \bibinfo {pages} {012001} (\bibinfo {year}
  {2022})}\BibitemShut {NoStop}%
\bibitem [{\citenamefont {Houston}\ and\ \citenamefont
  {Alexander}(2023)}]{Alexander2023}%
  \BibitemOpen
  \bibfield  {author} {\bibinfo {author} {\bibfnamefont {A.~J.~H.}\
  \bibnamefont {Houston}}\ and\ \bibinfo {author} {\bibfnamefont {G.~P.}\
  \bibnamefont {Alexander}},\ }\bibfield  {title} {\bibinfo {title} {Colloids
  in two-dimensional active nematics: conformal cogs and controllable
  spontaneous rotation},\ }\href@noop {} {\bibfield  {journal} {\bibinfo
  {journal} {New Journal of Physics}\ }\textbf {\bibinfo {volume} {25}},\
  \bibinfo {pages} {123006} (\bibinfo {year} {2023})}\BibitemShut {NoStop}%
\bibitem [{\citenamefont {Neville}\ \emph {et~al.}(2024)\citenamefont
  {Neville}, \citenamefont {Eggers},\ and\ \citenamefont
  {Liverpool}}]{Neville2024}%
  \BibitemOpen
  \bibfield  {author} {\bibinfo {author} {\bibfnamefont {L.}~\bibnamefont
  {Neville}}, \bibinfo {author} {\bibfnamefont {J.}~\bibnamefont {Eggers}},\
  and\ \bibinfo {author} {\bibfnamefont {T.~B.}\ \bibnamefont {Liverpool}},\
  }\bibfield  {title} {\bibinfo {title} {Controlling wall--particle
  interactions with activity},\ }\href@noop {} {\bibfield  {journal} {\bibinfo
  {journal} {Soft Matter}\ }\textbf {\bibinfo {volume} {20}},\ \bibinfo {pages}
  {8395} (\bibinfo {year} {2024})}\BibitemShut {NoStop}%
\bibitem [{\citenamefont {Shendruk}\ and\ \citenamefont
  {Yeomans}(2015)}]{Shendruk2015SoftMatter-NMPCD}%
  \BibitemOpen
  \bibfield  {author} {\bibinfo {author} {\bibfnamefont {T.~N.}\ \bibnamefont
  {Shendruk}}\ and\ \bibinfo {author} {\bibfnamefont {J.~M.}\ \bibnamefont
  {Yeomans}},\ }\bibfield  {title} {\bibinfo {title} {Multi-particle collision
  dynamics algorithm for nematic fluids},\ }\href@noop {} {\bibfield  {journal}
  {\bibinfo  {journal} {Soft Matter}\ }\textbf {\bibinfo {volume} {11}},\
  \bibinfo {pages} {5101} (\bibinfo {year} {2015})}\BibitemShut {NoStop}%
\bibitem [{\citenamefont {Valei}\ \emph {et~al.}(2025)\citenamefont {Valei},
  \citenamefont {Wamsler}, \citenamefont {Parker}, \citenamefont {Obara},
  \citenamefont {Klotz},\ and\ \citenamefont
  {Shendruk}}]{valei2025-PassivePolymerMPCD}%
  \BibitemOpen
  \bibfield  {author} {\bibinfo {author} {\bibfnamefont {Z.~K.}\ \bibnamefont
  {Valei}}, \bibinfo {author} {\bibfnamefont {K.}~\bibnamefont {Wamsler}},
  \bibinfo {author} {\bibfnamefont {A.~J.}\ \bibnamefont {Parker}}, \bibinfo
  {author} {\bibfnamefont {T.~A.}\ \bibnamefont {Obara}}, \bibinfo {author}
  {\bibfnamefont {A.~R.}\ \bibnamefont {Klotz}},\ and\ \bibinfo {author}
  {\bibfnamefont {T.~N.}\ \bibnamefont {Shendruk}},\ }\bibfield  {title}
  {\bibinfo {title} {Dynamics of polymers in coarse-grained nematic solvents},\
  }\href@noop {} {\bibfield  {journal} {\bibinfo  {journal} {Soft Matter}\
  }\textbf {\bibinfo {volume} {21}},\ \bibinfo {pages} {361} (\bibinfo {year}
  {2025})}\BibitemShut {NoStop}%
\bibitem [{\citenamefont {Hemingway}\ \emph {et~al.}(2016)\citenamefont
  {Hemingway}, \citenamefont {Mishra}, \citenamefont {Marchetti},\ and\
  \citenamefont {Fielding}}]{Hemingway2016SoftMatter}%
  \BibitemOpen
  \bibfield  {author} {\bibinfo {author} {\bibfnamefont {E.~J.}\ \bibnamefont
  {Hemingway}}, \bibinfo {author} {\bibfnamefont {P.}~\bibnamefont {Mishra}},
  \bibinfo {author} {\bibfnamefont {M.~C.}\ \bibnamefont {Marchetti}},\ and\
  \bibinfo {author} {\bibfnamefont {S.~M.}\ \bibnamefont {Fielding}},\
  }\bibfield  {title} {\bibinfo {title} {Correlation lengths in hydrodynamic
  models of active nematics},\ }\href@noop {} {\bibfield  {journal} {\bibinfo
  {journal} {Soft Matter}\ }\textbf {\bibinfo {volume} {12}},\ \bibinfo {pages}
  {7943} (\bibinfo {year} {2016})}\BibitemShut {NoStop}%
\bibitem [{\citenamefont {Giomi}\ \emph
  {et~al.}(2014{\natexlab{b}})\citenamefont {Giomi}, \citenamefont {Bowick},
  \citenamefont {Mishra}, \citenamefont {Sknepnek},\ and\ \citenamefont
  {Cristina~Marchetti}}]{Giomi2014PhilTransactionsA}%
  \BibitemOpen
  \bibfield  {author} {\bibinfo {author} {\bibfnamefont {L.}~\bibnamefont
  {Giomi}}, \bibinfo {author} {\bibfnamefont {M.~J.}\ \bibnamefont {Bowick}},
  \bibinfo {author} {\bibfnamefont {P.}~\bibnamefont {Mishra}}, \bibinfo
  {author} {\bibfnamefont {R.}~\bibnamefont {Sknepnek}},\ and\ \bibinfo
  {author} {\bibfnamefont {M.}~\bibnamefont {Cristina~Marchetti}},\ }\bibfield
  {title} {\bibinfo {title} {Defect dynamics in active nematics},\ }\href@noop
  {} {\bibfield  {journal} {\bibinfo  {journal} {Philosophical Transactions of
  the Royal Society A}\ }\textbf {\bibinfo {volume} {372}},\ \bibinfo {pages}
  {20130365} (\bibinfo {year} {2014}{\natexlab{b}})}\BibitemShut {NoStop}%
\bibitem [{\citenamefont {Thampi}\ \emph {et~al.}(2014)\citenamefont {Thampi},
  \citenamefont {Golestanian},\ and\ \citenamefont {Yeomans}}]{Thampi2014EPL}%
  \BibitemOpen
  \bibfield  {author} {\bibinfo {author} {\bibfnamefont {S.~P.}\ \bibnamefont
  {Thampi}}, \bibinfo {author} {\bibfnamefont {R.}~\bibnamefont
  {Golestanian}},\ and\ \bibinfo {author} {\bibfnamefont {J.~M.}\ \bibnamefont
  {Yeomans}},\ }\bibfield  {title} {\bibinfo {title} {Instabilities and
  topological defects in active nematics},\ }\href@noop {} {\bibfield
  {journal} {\bibinfo  {journal} {Europhysics Letters}\ }\textbf {\bibinfo
  {volume} {105}},\ \bibinfo {pages} {18001} (\bibinfo {year}
  {2014})}\BibitemShut {NoStop}%
\bibitem [{\citenamefont {Head}\ \emph
  {et~al.}(2024{\natexlab{b}})\citenamefont {Head}, \citenamefont {Dor{\'e}},
  \citenamefont {Keogh}, \citenamefont {Bonn}, \citenamefont {Negro},
  \citenamefont {Marenduzzo}, \citenamefont {Doostmohammadi}, \citenamefont
  {Thijssen}, \citenamefont {L{\'o}pez-Le{\'o}n},\ and\ \citenamefont
  {Shendruk}}]{Head2024-DTensor}%
  \BibitemOpen
  \bibfield  {author} {\bibinfo {author} {\bibfnamefont {L.~C.}\ \bibnamefont
  {Head}}, \bibinfo {author} {\bibfnamefont {C.}~\bibnamefont {Dor{\'e}}},
  \bibinfo {author} {\bibfnamefont {R.~R.}\ \bibnamefont {Keogh}}, \bibinfo
  {author} {\bibfnamefont {L.}~\bibnamefont {Bonn}}, \bibinfo {author}
  {\bibfnamefont {G.}~\bibnamefont {Negro}}, \bibinfo {author} {\bibfnamefont
  {D.}~\bibnamefont {Marenduzzo}}, \bibinfo {author} {\bibfnamefont
  {A.}~\bibnamefont {Doostmohammadi}}, \bibinfo {author} {\bibfnamefont
  {K.}~\bibnamefont {Thijssen}}, \bibinfo {author} {\bibfnamefont
  {T.}~\bibnamefont {L{\'o}pez-Le{\'o}n}},\ and\ \bibinfo {author}
  {\bibfnamefont {T.~N.}\ \bibnamefont {Shendruk}},\ }\bibfield  {title}
  {\bibinfo {title} {Spontaneous self-constraint in active nematic flows},\
  }\href@noop {} {\bibfield  {journal} {\bibinfo  {journal} {Nature Physics}\ }
  (\bibinfo {year} {2024}{\natexlab{b}})}\BibitemShut {NoStop}%
\bibitem [{\citenamefont {Musevic}\ \emph {et~al.}(2006)\citenamefont
  {Musevic}, \citenamefont {Skarabot}, \citenamefont {Tkalec}, \citenamefont
  {Ravnik},\ and\ \citenamefont {Zumer}}]{Musevic2006}%
  \BibitemOpen
  \bibfield  {author} {\bibinfo {author} {\bibfnamefont {I.}~\bibnamefont
  {Musevic}}, \bibinfo {author} {\bibfnamefont {M.}~\bibnamefont {Skarabot}},
  \bibinfo {author} {\bibfnamefont {U.}~\bibnamefont {Tkalec}}, \bibinfo
  {author} {\bibfnamefont {M.}~\bibnamefont {Ravnik}},\ and\ \bibinfo {author}
  {\bibfnamefont {S.}~\bibnamefont {Zumer}},\ }\bibfield  {title} {\bibinfo
  {title} {Two-dimensional nematic colloidal crystals self-assembled by
  topological defects},\ }\href@noop {} {\bibfield  {journal} {\bibinfo
  {journal} {Science}\ }\textbf {\bibinfo {volume} {313}},\ \bibinfo {pages}
  {954} (\bibinfo {year} {2006})}\BibitemShut {NoStop}%
\bibitem [{\citenamefont {{\v{S}}karabot}\ \emph {et~al.}(2008)\citenamefont
  {{\v{S}}karabot}, \citenamefont {Ravnik}, \citenamefont {{\v{Z}}umer},
  \citenamefont {Tkalec}, \citenamefont {Poberaj}, \citenamefont
  {Babi{\v{c}}},\ and\ \citenamefont {Mu{\v{s}}evi{\v{c}}}}]{Skarabot2008}%
  \BibitemOpen
  \bibfield  {author} {\bibinfo {author} {\bibfnamefont {M.}~\bibnamefont
  {{\v{S}}karabot}}, \bibinfo {author} {\bibfnamefont {M.}~\bibnamefont
  {Ravnik}}, \bibinfo {author} {\bibfnamefont {S.}~\bibnamefont {{\v{Z}}umer}},
  \bibinfo {author} {\bibfnamefont {U.}~\bibnamefont {Tkalec}}, \bibinfo
  {author} {\bibfnamefont {I.}~\bibnamefont {Poberaj}}, \bibinfo {author}
  {\bibfnamefont {D.}~\bibnamefont {Babi{\v{c}}}},\ and\ \bibinfo {author}
  {\bibfnamefont {I.}~\bibnamefont {Mu{\v{s}}evi{\v{c}}}},\ }\bibfield  {title}
  {\bibinfo {title} {Hierarchical self-assembly of nematic colloidal
  superstructures},\ }\href@noop {} {\bibfield  {journal} {\bibinfo  {journal}
  {Physical Review E}\ }\textbf {\bibinfo {volume} {77}},\ \bibinfo {pages}
  {061706} (\bibinfo {year} {2008})}\BibitemShut {NoStop}%
\bibitem [{\citenamefont {Ray}\ \emph {et~al.}(2023)\citenamefont {Ray},
  \citenamefont {Zhang},\ and\ \citenamefont {Dogic}}]{Sattvic2023}%
  \BibitemOpen
  \bibfield  {author} {\bibinfo {author} {\bibfnamefont {S.}~\bibnamefont
  {Ray}}, \bibinfo {author} {\bibfnamefont {J.}~\bibnamefont {Zhang}},\ and\
  \bibinfo {author} {\bibfnamefont {Z.}~\bibnamefont {Dogic}},\ }\bibfield
  {title} {\bibinfo {title} {Rectified rotational dynamics of mobile inclusions
  in two-dimensional active nematics},\ }\href@noop {} {\bibfield  {journal}
  {\bibinfo  {journal} {Physical Review Letters}\ }\textbf {\bibinfo {volume}
  {130}},\ \bibinfo {pages} {238301} (\bibinfo {year} {2023})}\BibitemShut
  {NoStop}%
\bibitem [{\citenamefont {Guzm{\'a}n-Lastra}\ \emph {et~al.}(2021)\citenamefont
  {Guzm{\'a}n-Lastra}, \citenamefont {L{\"o}wen},\ and\ \citenamefont
  {Mathijssen}}]{Guzman2021}%
  \BibitemOpen
  \bibfield  {author} {\bibinfo {author} {\bibfnamefont {F.}~\bibnamefont
  {Guzm{\'a}n-Lastra}}, \bibinfo {author} {\bibfnamefont {H.}~\bibnamefont
  {L{\"o}wen}},\ and\ \bibinfo {author} {\bibfnamefont {A.~J.}\ \bibnamefont
  {Mathijssen}},\ }\bibfield  {title} {\bibinfo {title} {Active carpets drive
  non-equilibrium diffusion and enhanced molecular fluxes},\ }\href@noop {}
  {\bibfield  {journal} {\bibinfo  {journal} {Nature Communications}\ }\textbf
  {\bibinfo {volume} {12}},\ \bibinfo {pages} {1906} (\bibinfo {year}
  {2021})}\BibitemShut {NoStop}%
\end{thebibliography}%
